\documentclass[preprint,aps]{revtex4}
\pdfoutput=1

\usepackage{graphicx}
\usepackage{dcolumn}
\usepackage{bm}
\usepackage{color}

\begin{document}


\title{Signatures of Incipient Jamming in Collisional Hopper Flows}

\author{Shubha Tewari}
\email{shubha.tewari@wne.edu}
\affiliation{Department of Physical and Biological Sciences, 
Western New England University, 1215 Wilbraham Road, Springfield, MA 01119}
\author{Michal Dichter}
\author{Bulbul Chakraborty}
\affiliation{Martin Fisher School of Physics, Brandeis University,
Mailstop 057, Waltham, MA 02454-9110 }

\date{\today}

\begin{abstract}

Many disordered systems experience a transition from a fluid-like state to a solid-like state following a sudden arrest in dynamics called jamming. In contrast to jamming in spatially homogeneous systems, jamming in hoppers occurs under extremely inhomogeneous conditions as the gravity-driven flow of grains enclosed by rigid walls converges towards a small opening.  In this work, we study velocity fluctuations in a collisional flow  near jamming using event-driven simulations. The average flow in a hopper geometry is known to have strong gradients, especially near the walls and the orifice.  We find, in addition, a spatially heterogeneous distribution of fluctuations, most striking in the velocity autocorrelation relaxation times. At high flow rates, the flow at the center has lower kinetic temperatures and longer autocorrelation times than at the boundary. Remarkably, however, this trend reverses itself as the flow rate slows, with fluctuations relaxing more slowly at the boundaries though the kinetic temperatures remain high in that region. The slowing down of the dynamics is accompanied by increasing non-Gaussianity in the velocity distributions, which also have large spatial variations.   
\end{abstract}

\maketitle

\section{\label{sec:level1}Introduction}

Granular flows can clog unpredictably, often with catastrophic consequences as seen in grain silo failures\cite{grain_silo}.  This phenomenon is an example of jamming, where a disordered system undergoes a sudden arrest in dynamics, leading to a transition from a fluid-like to a solid-like state\cite{liunagel}.   
Unlike the jams that develop in spatially homogeneous systems, however, jams in hopper or silo flows occur under extremely inhomogeneous
conditions as a gravity-driven flow enclosed by rigid walls converges towards an orifice where stable arches can form\cite{to_lai,garcimartin:2010pre,tang_behringer,thomas_durian,zuriguel:2011prl}. Hopper flows are self-organized, being neither pressure nor density controlled, with a steady state that develops from an interplay between gravitational forcing, constraints due to the walls, and the interactions between grains\cite{jnb_rmp}. The flow is also driven by body forces, and the boundary is thus not a source of energy.  Instead, the boundary imposes constraints on the flow volume, is a source of frictional interaction, and randomizes the flow through grain-wall collisions. The dynamical principles that lead to jamming in such self-regulated flows are not well understood.

Experimental investigations of two-dimensional (2D) hopper flows show that these flows remain collisional near jamming\cite{gardel}, and show large force fluctuations\cite{longhi} and transient arch formation\cite{tang_behringer}.  
In previous studies\cite{ally_epl04,ally_pre}, we have shown that collisional flows with no extended contacts develop transient force chains that are are sustained by correlated collisions. The origin of these correlations is the inelasticity of the collisions, and the stress is purely due to momentum transfer.  Focusing on the homogeneous flow in the bulk of the hopper, we established the emergence of growing length and time scales as the flow approached jamming\cite{Ally_07,shub_pre}.  Flows in hoppers, however, are characterized by strong gradients, such as a shear-layer at the walls\cite{pouliquen_gutfraind,losert}, and rapid flow regions near the orifice.  In addition to these heterogeneities in the average flow profile, the fluctuations in the flow have well-defined spatial structure, such as transient arches\cite{tang_behringer}.  In this work, we focus on the dynamics in regions with strong gradients and their vital role in controlling the approach to jamming.
 
Using event-driven simulations of a purely collisional, gravity-driven flow, we monitor the temporal variation of the flow in boxes that are a few grain diameters in size.  
We observe that the approach to jamming in the collisional non-equilibrium steady state (NESS) is signaled by the flow becoming intermittent with time intervals spanning many collisions in which the flow alternately slows down and speeds up. During some of these intervals, the flow is slow (fast) throughout the hopper but in other instances, the slowing down (speeding up) is localized in the bulk or near the boundaries. These fluctuations are particularly strong in hoppers with small openings\cite{gardel}. 
In addition to the marked gradients in the flow field, the fluctuation-statistics of the velocity show large variations both along and transverse to the flow direction. More importantly, the spatial patterns evolve with flow rate and can change quite dramatically as the flow approaches jamming.  In particular, we observe a flowing regime, and a pre-jamming regime distinguished by the spatial behavior of the velocity relaxation. In a rare jamming event that spans millions of collisions,  we see evidence of the boundaries  frustrating the tendency to develop a uniform, steady flow in response to the gravitational forcing.


\section{Description of simulation}
Our event-driven simulation is based on a particle dynamics model used by Denniston and Li \cite{denniston_li} and is described in detail in earlier papers\cite{ally_epl04,ally_pre}. The system consists of 1,000 non-deformable disks falling under gravity in a 2D rectangular hopper with a tapered base (see Fig. \ref {fig:vfield}). The disks are bidisperse, with diameters $d$ and $1.2d$; collisions between particles are inelastic but frictionless, so momentum transfer occurs along the center-to-center vector of colliding particles. The transformation in relative velocity between colliding particles $i$ and $j$ is defined in terms of a coefficient of restitution $\mu$:
\begin{equation}
(\bm{u}_{j}^{\prime} - \bm{u}_{i}^{\prime}).\hat{q} = -\mu (\bm{u}_j - \bm{u}_i).\hat{q}
\end{equation}
where $\bm{u}$ , $\bm{u}^{\prime}$ are the particle velocities before and after the collision, and $\hat{q}$ is a unit vector along the center-to-center direction. All inter-particle collisions become elastic below a relative velocity threshold $u_{\text{cut}}$ to avoid inelastic collapse\cite{bennaim_redner}. Particle-wall collisions are inelastic, with frictional drag at the walls modeled by a tangential coefficient of restitution $\mu_{\text{wall}}$, allowing the flow of grains through the outlet at the bottom of the system to reach a steady-state. Particles exiting the system at the base are reintroduced at random lateral positions at the top of the hopper. 

All collisions in this event-driven simulation are instantaneous, and particles cannot form temporally extended contacts with one another. 
The simulation moves forward from collision to collision, and instantaneous snapshots of the system are reconstructed from the state of the system before and after the relevant collisions to determine particle positions and velocities at equally spaced time intervals. The overall flow rate is set by the particle-wall friction coefficient and the size of the opening and varies linearly with opening size, as we show in Figure~\ref{fig:vfield}. The mass of the smaller grains and the acceleration of gravity $g$ are set to 1, and all lengths are expressed in units of the smaller particle diameter $d$.  In these units, the length and width of the rectangular region are 76.5 and 20 respectively, the angle of the tapered base is 45 degrees, and the size of the opening at the base is varied between 10 and 4.5. The simulation parameters for the collisions are $\mu = 0.8$, $\mu_{\text{wall}} = 0.5$,  and $u_{\text{cut}}= 10^{-3}$. The time taken for free fall of a particle through its own diameter is $\sqrt{2}$, and the average time between collisions for a given particle is on the order of $10^{-3}$. 

The development of extended contacts is, of course, important for sustaining a jam.  It is not clear, however, if the approach to jamming is qualitatively altered by the presence of extended contacts.   Frequent collisions can mimic extended contacts\cite{ally_epl04,ally_pre}, and one can construct an effective potential for hard spheres  by coarse-graining configurations over time intervals large compared to times between collisions\cite{brito_wyart,donev_torquato}.   In this paper, we discuss signatures of an approaching jam in collisional flows, assuming that a suitable coarse-graining can map these on to flows that develop extended contacts.  The majority of our results are from the intermittent flow regime where we observe clearly defined intervals of slow and fast flows.  In addition, we have studied one jam that lasts for $10^7$ collisions, and characterized the structure of the flow during its creation and breakup. A permanent jam cannot be created in our model of collisional flow since we prevent inelastic collapse\cite{bennaim_redner}.

\section{Results}
\begin{figure}
\includegraphics[width=\textwidth]{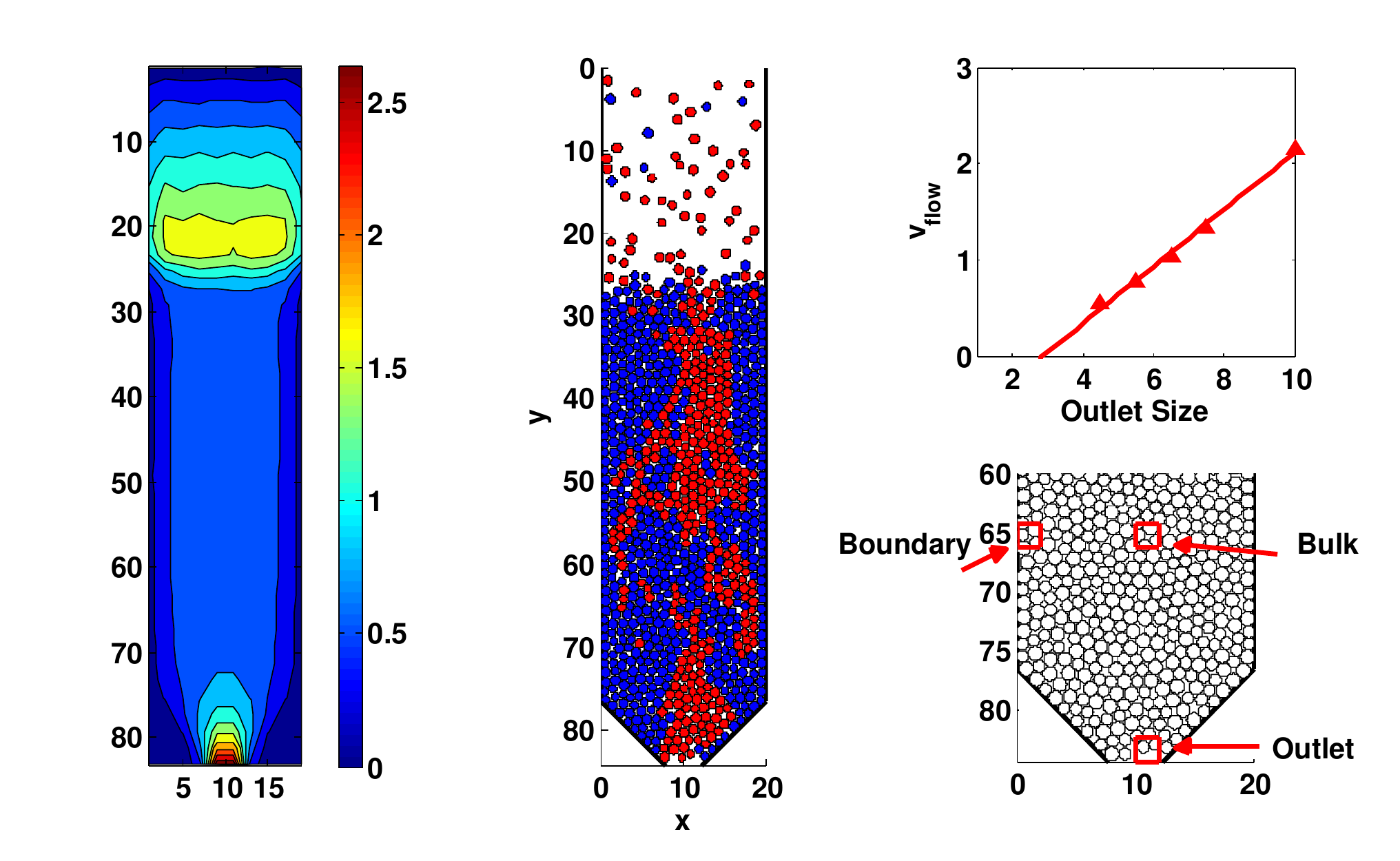}
\caption{\label{fig:vfield} (Color Online) Left: Spatial contour map of $\mathcal{V}_y(\bm{r})$, illustrating a gradient-free region in the center, a shear layer bordering the walls, and a region of rapid flow near the orifice. Center: A snapshot of all the particle positions at a given instant in time, showing fast particles with $u_{i,y}(t) \ge \mathcal{V}_y(t)$ (red) and slow particles with $u_{i,y} <  \mathcal{V}_y(t)$ (blue).  Upper right: Mean flow rate, $v_{\text{flow}} = \int dt  {\mathcal{V}}_y ({t})$, measured in simulation units, shown as a function of the outlet size measured in units of $d$. Lower right: Locations (\{x,y\}) of three boxes that exemplify three very different regions of the hopper.}
\end{figure}

To study fluctuations, we construct a velocity field, $\bm{v}(\bm{r},t)$ by coarse graining 
individual particle velocities $\bm{u}_{i}$ over a box of size $2d \times 2d$:
\begin{equation}
\bm{v}(\bm{r},t) = {1\over N}\sum_{j=1}^{N} \bm{u}_{j}(t)
\end{equation} 
where the sum is over the $N$ particles
whose centers lie inside a box centered at $\bm{r} =(x,y)$ at time $t$. The entire hopper is subdivided into non-overlapping boxes,  starting at the base of the hopper and proceeding vertically upwards.  We employ two different types of averaging to measure fluctuations. A time-averaged flow-field, $\bm{\mathcal{V}} (\bm{r}) = \int dt \bm{v}(\bm{r},t)$ may be used to measure temporal fluctuations in each box; and a spatially averaged flow-field, $\bm{\mathcal{V}} (t) = \int d\bm{r}  \bm{v}(\bm{r},t)$ may be used to measure instantaneous variations of the velocity. Both measures are illustrated in Figure~\ref{fig:vfield}, which shows the spatio-temporal heterogeneity of the velocity field. In most of the analysis we describe in this paper, we use the box-specific definition, $\bm{\mathcal{V}} (\bm{r})$.

The leftmost panel of Figure~\ref{fig:vfield} shows the spatial pattern in $\mathcal{V}_y(\bm{r})$.  Particles enter the hopper at the top and large gradients are seen in the direction of the flow.   This region transitions into the ``bulk'' where  $\mathcal{V}_y(\bm{r})$ is gradient-free in the direction of the flow, but a marked interface with a large gradient separates the bulk flow from the shear layer along the vertical wall.  This bulk flow gives way to a flow with strong gradients, transverse to and along the flow direction, close to the orifice.   The length scales associated with the gradients of $\mathcal{V}_y(\bm{r})$ are not well understood.  The width of the shear layer has been related to a stress-activated mechanism\cite{pouliquen_gutfraind}, and we are currently investigating the relation between stress fluctuations and velocity gradients in our simulations\cite{st_bc_inprep}.

The center panel of Figure~\ref{fig:vfield} shows an example of instantaneous variations in the velocity field.  Tracking the velocities of individual particles, $\bm{u}_i (t)$, we label them as fast ($u_{i,y} (t) \ge \mathcal{V}_y ({t})$) and slow ($u_{i,y} (t) < \mathcal{V}_y ({t})$).   The snapshot shows large, vertically-extended clusters of fast moving particles, demonstrating non-trivial spatio-temporal correlations in the flow.

The average flow field, $\mathcal{V}_y(\bm{r})$ identifies three interesting regions of the flow, if we set aside the region at the top of the hopper where particles are entering.  To analyze fluctuations in these regions, we choose three representative boxes: a box at the orifice, a box in the bulk flow region, and one in the shear layer; these three are shown in the lower right panel of Figure~\ref{fig:vfield}.  In addition, to better characterize the spatial variation of the fluctuations, we study their statistics in boxes along the central vertical column, and a horizontal row running through the bulk flow region.  These two cuts span the regions of strong gradients in $\mathcal{V}_y(\bm{r})$.   The mean flow rate, $v_{\text{flow}} = \int dt  {\mathcal{V}}_y ({t})$, is controlled by the size of the hopper opening, and as the upper right panel of Figure~\ref{fig:vfield} shows, varies linearly with the outlet width.

\subsection{Velocity Autocorrelations}

\begin{figure}
\hspace{-0.7in}
\includegraphics[width=0.8\textwidth]{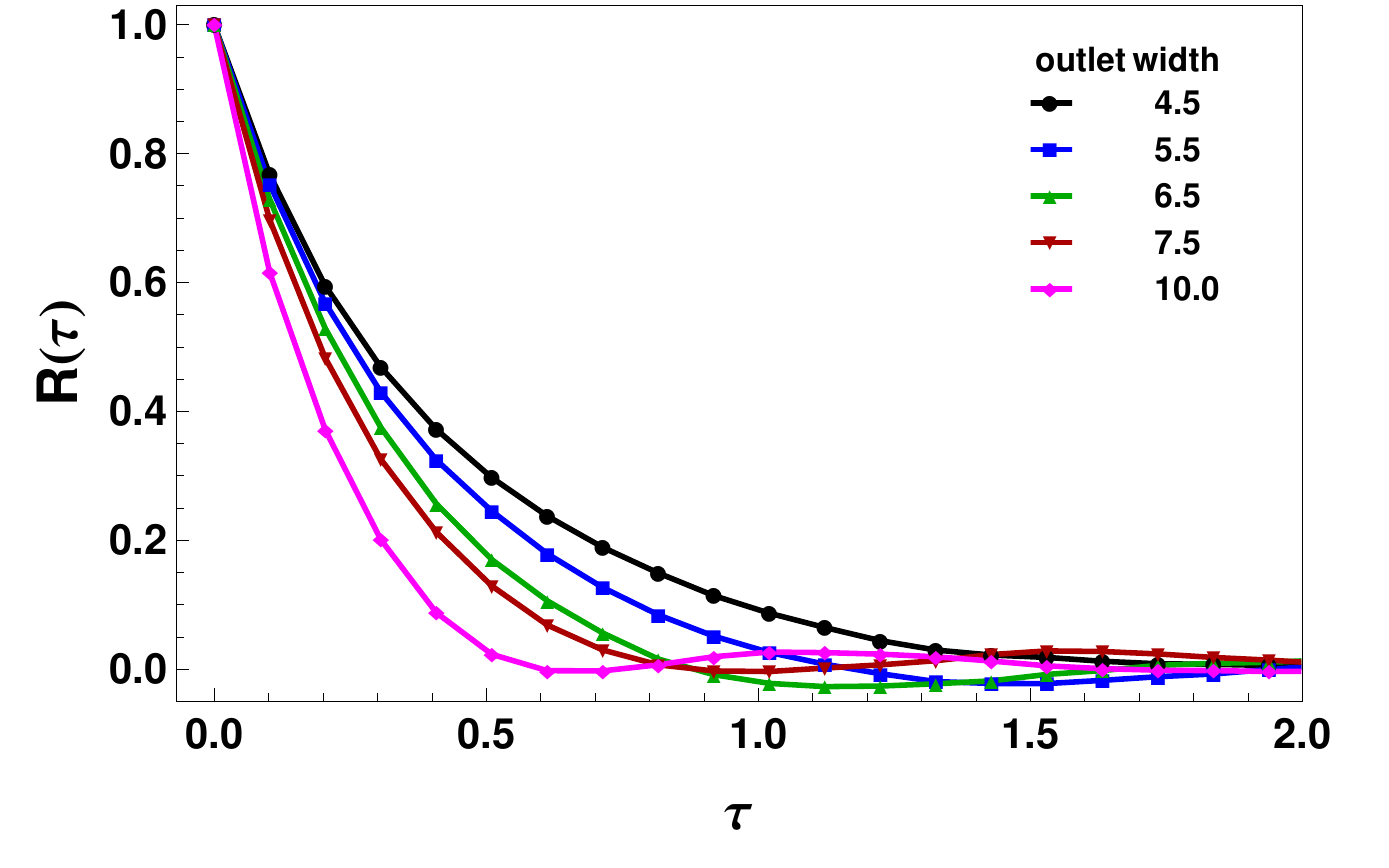}
\includegraphics[width=0.4\textwidth]{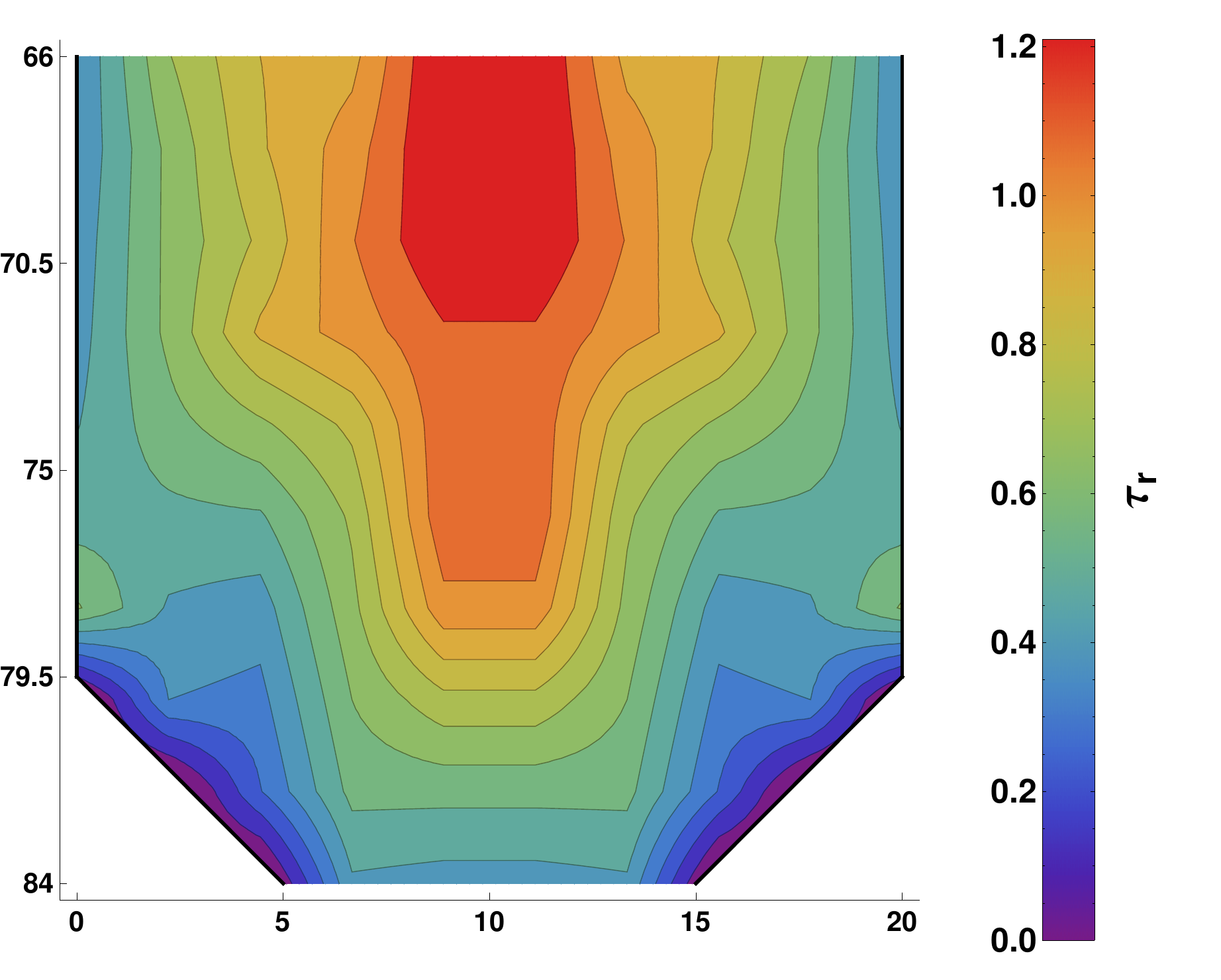}
\includegraphics[width=0.4\textwidth]{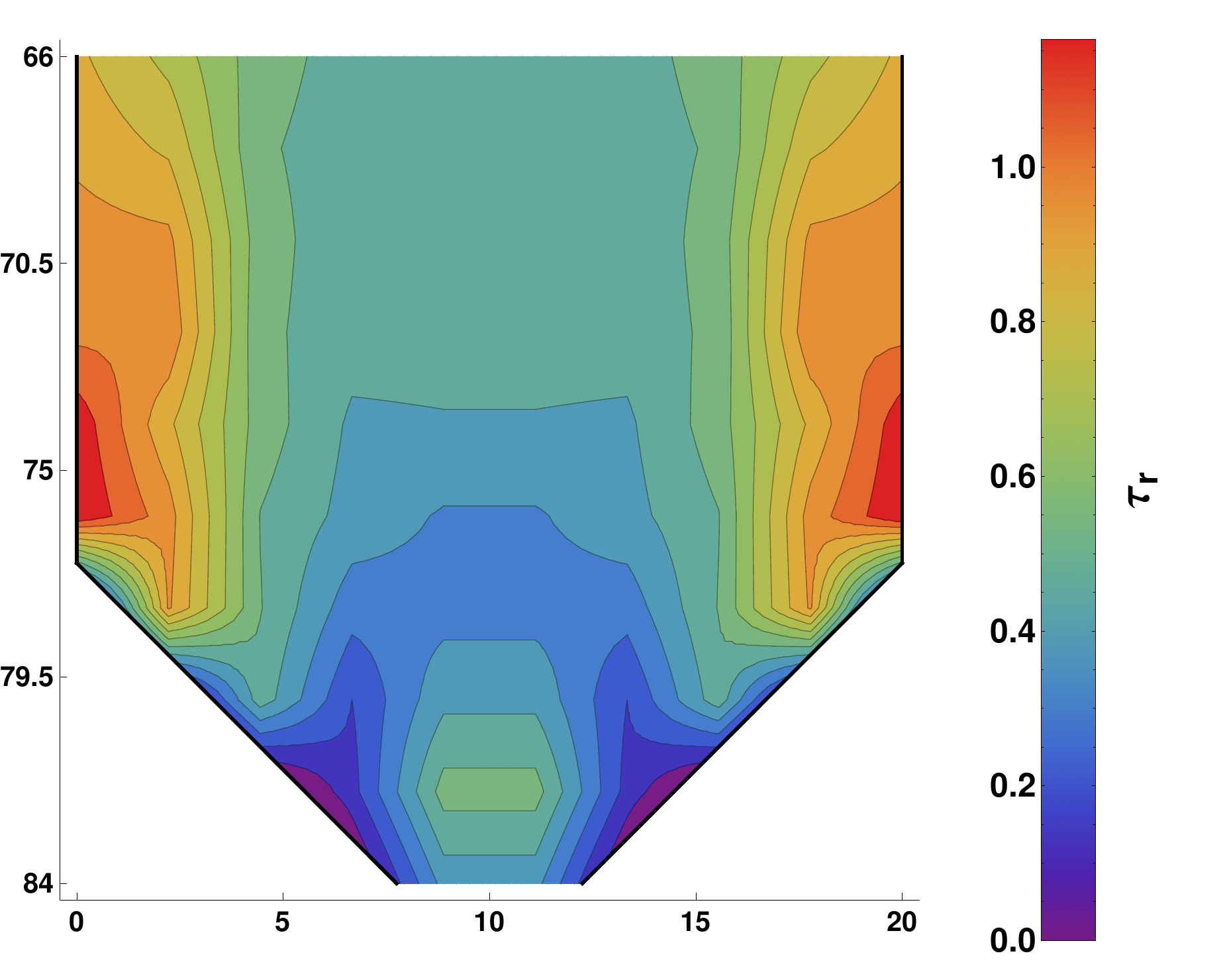}
\caption{\label{fig:vautocorr} (Color Online) Upper panel: $C(\bf{r}, t)$ measured at the boundary box at different outlet openings.  Lower panel:  Spatial variation of $\tau(\bf{r})$ at outlet size 10.0 (left), which was the fastest flow rate simulated, and at outlet size 4.5 (right), which was the slowest flow rate simulated.}
\end{figure}

The approach to jamming is marked by large temporal variations in $v_y({\bf{r}},t)$.  As we show in this section,  the timescales associated with velocity fluctuations provide the strongest signal of an incipient jam.
We study the statistics of the temporal fluctuations through the velocity autocorrelation function, $C({\bf{r}}, t) = \langle v_y({\bf{r}},t) v_y({\bf{r}},0) \rangle$.  
As an example of the behavior of $C({\bf{r}}, t)$, the upper panel of Figure~\ref{fig:vautocorr} shows its variation with flow rate in the shear-layer as measured in the box closest to the vertical wall.   If we measure the autocorrelation time by $C({\bf{r}}, \tau({\bf{r}})) = 0.1$, then $\tau$  in this box increases by a factor of 2 as the outlet size changes from 10 to 4.5.  The spatial variation of the relaxation times of the velocity fluctuations is represented by the field $\tau({\bf{r}})$, which is shown in  Figure~\ref{fig:vautocorr} for the fastest and the slowest flow rates.  The plots are symmetrized about the central vertical line, so there is meaningful information only in half of each plot.

The lower left panel of Figure ~\ref{fig:vautocorr} illustrates the pattern of $\tau(\bf{r})$ at an opening of 10.0.      
This fast-flowing NESS  is characterized by regions of fast relaxation close to the boundary, and increasingly slow relaxations as one moves towards the center of the hopper.  At slower flow rates,  a clear reversal in trend is evident from the lower right panel of Figure~\ref{fig:vautocorr}.  At opening size 4.5, $\tau(\bf{r})$ is longest at the walls, and shortest at the center.  The change in the pattern of relaxation times signals a qualitative change in the behavior of the flow as it approaches jamming, and is the most significant dynamical signature that we have observed in our simulations.   

The near-jamming pattern poses a puzzle if we view the NESS as a fluid characterized by a granular  temperature\cite{gran_temp_ref}.   The kinetic granular temperature is proportional to the variance of the velocity distribution.  As shown in our earlier paper\cite{shub_pre}, and detailed later in Table~\ref{mean_deviation}, the kinetic temperature increases from the center to the shear layer with a sharp peak very close to the vertical wall.  This trend does not change with flow rate.  Near jamming, therefore, velocity fluctuations relax more slowly in regions with higher kinetic temperature.  This behavior is at odds with our intuition, based on thermal fluids, of slower relaxations at lower temperatures.  For the fast flows, the hopper-NESS does show the expected behavior.  We conclude that there is a pre-jamming regime in collisional hopper flows where the NESS behaves neither like a fluid nor a solid.

If we {\it assume} that the Green-Kubo relations\cite{Green-Kubo} hold for the NESS in the hopper, then the autocorrelation time is related to the diffusion constant $1/\tau = D\propto k_B T/\eta$, where $T$ is the kinetic temperature and $\eta$ the viscosity.  The  viscosity is  related to fluctuations in the stress through another Green-Kubo relation.    At the fastest flow rate, we observe that $D$ is small at the center and increases towards the boundaries, consistent with the trend in $T$.  The fast-flowing NESS thus qualitatively resembles a fluid with temperature gradients.
For slow flows, $D$ is smaller at the boundaries than the bulk.  One possible explanation is an anomalously large viscosity $\eta$ near the boundaries, which just trades one puzzle for another since viscosity should also decrease with increasing temperature.  We do not understand the origin of the slow-relaxation pattern close to jamming.   A reasonable hypothesis is that boundaries frustrate the downward motion of the gravity-driven flow leading to slow relaxations.   In future work, we will study stress fluctuations\cite{st_bc_inprep} and their relationship to the velocity fluctuations in order to better understand the pre-jamming NESS.   

\begin{figure}
\vspace{-0.2in}
\includegraphics[width=\textwidth]{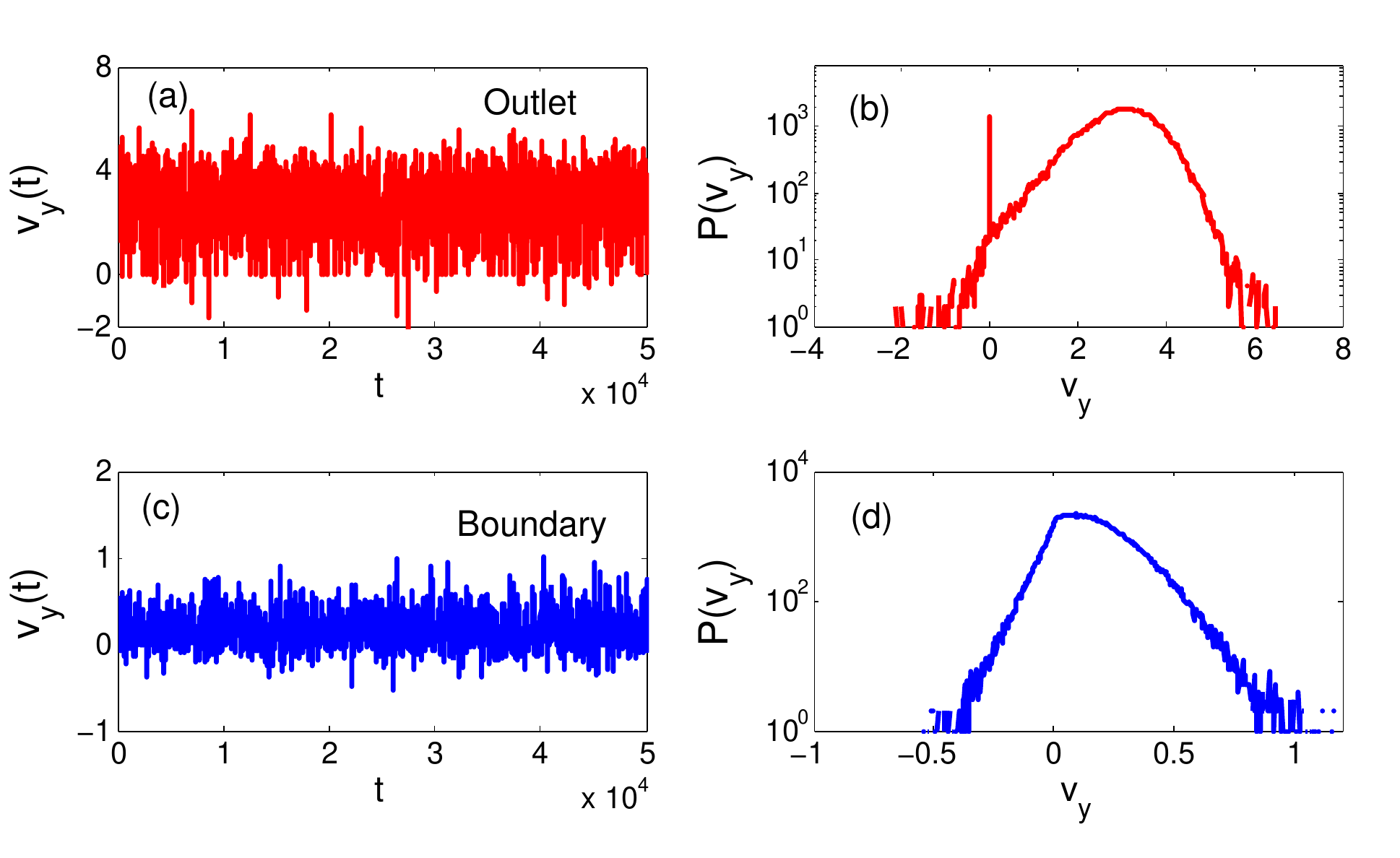}
\vspace{-0.5in}
\caption{\label{fig:timetrace} Illustrative examples of the time series of the velocity field, $v_y (\bm{r},t)$ at opening size 4.5.  (a) Time series measured in the box at the outlet, (b) the corresponding distribution; (c) Time series measured in the box at the vertical wall, (d) the corresponding distribution.}
\end{figure}

The pattern of relaxation times indicates that jamming is induced by the boundaries and propagates into the bulk.  As the flow slows, we also observe an increasing frequency of events with zero particles at the outlet.  These events are signaled by a peak at zero velocity in the velocity distribution function, seen in the upper right panel of Figure~\ref{fig:timetrace}. The percentage of these ``zero'' events remains small, varying from 0.14\% at opening 10.0 to 0.67\% at 4.5, as might be inferred from the time-trace of the velocity on the upper left side of this figure. The boxes in the shear layer also show a large number of zero and negative velocity events, as shown in the two lower panels of the same figure, consistent with the picture of a jam originating at the boundary. The shape of the distributions in these two boxes are distinct and will be discussed in Section C. In a future publication, we will analyze the negative fluctuations from the perspective of non-equilibrium fluctuation relations\cite{SSFR}.  

\subsection{First Passage Time Distributions}

In the pre-jamming regime, our analysis of the autocorrelation functions paints a picture of a collisional flow with fluctuations that do not fall within any familiar rubric of fluid flow. In this section, we analyze first passage processes in which the velocity transitions from below to above the average velocity, and those that go from above to below, to gain more insight into the nature of the intermittent flow that develops very close to jamming. It is well known\cite{bray_persistence} that for random processes, autocorrelation functions and first-passage probabilities provide complementary information. 

\begin{figure}
\includegraphics[scale=0.39]{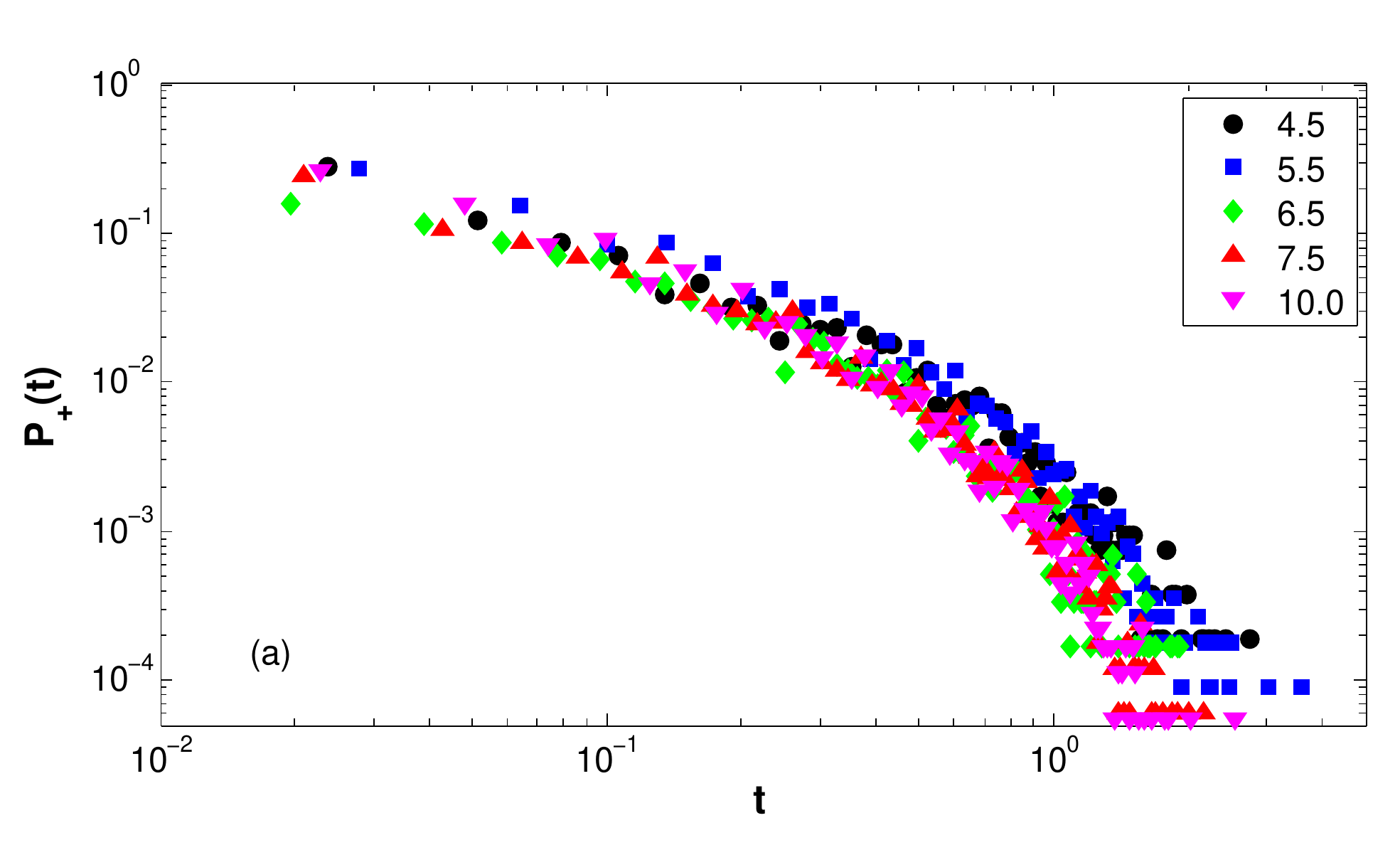}
\includegraphics[scale=0.39]{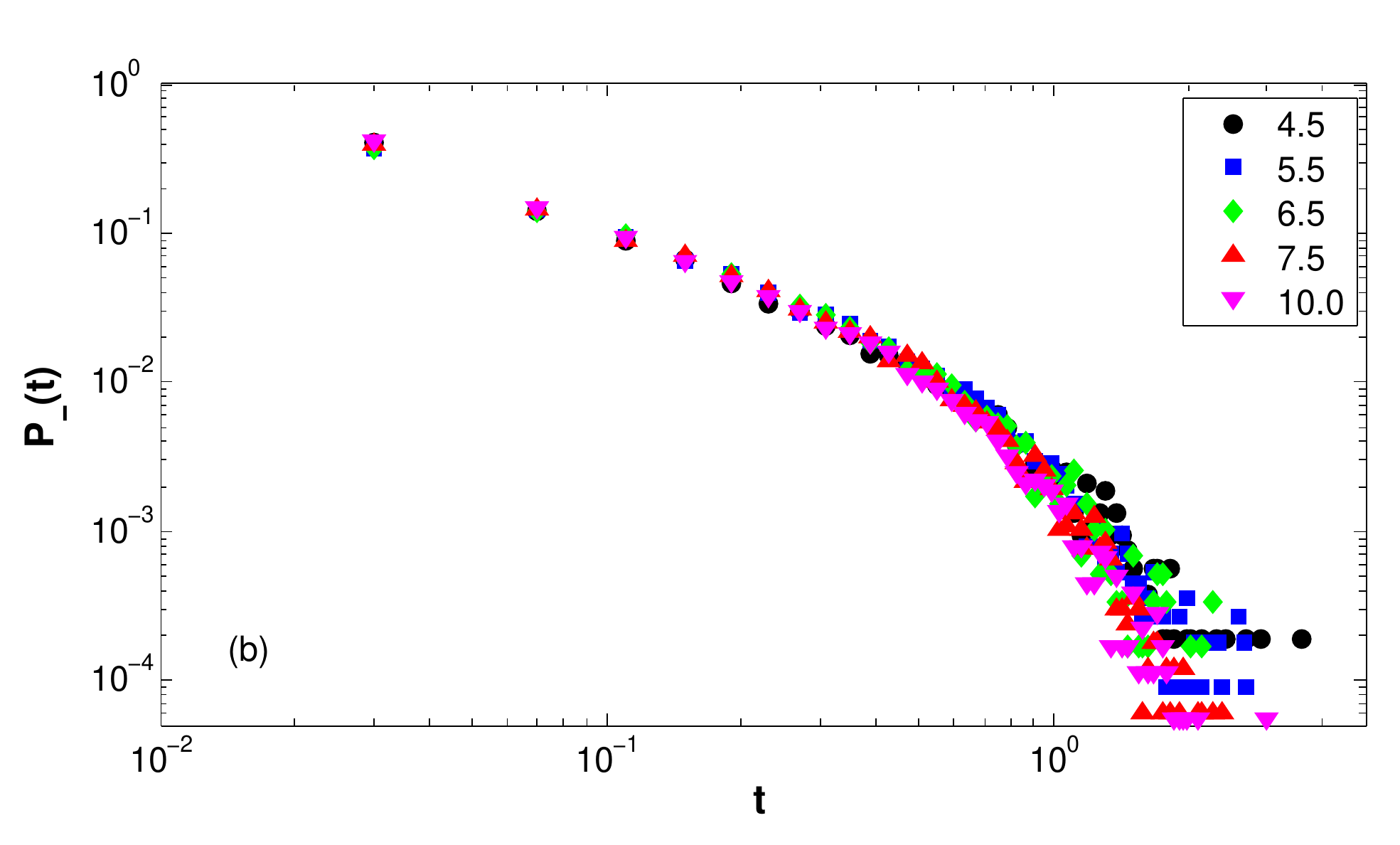}
\includegraphics[scale=0.39]{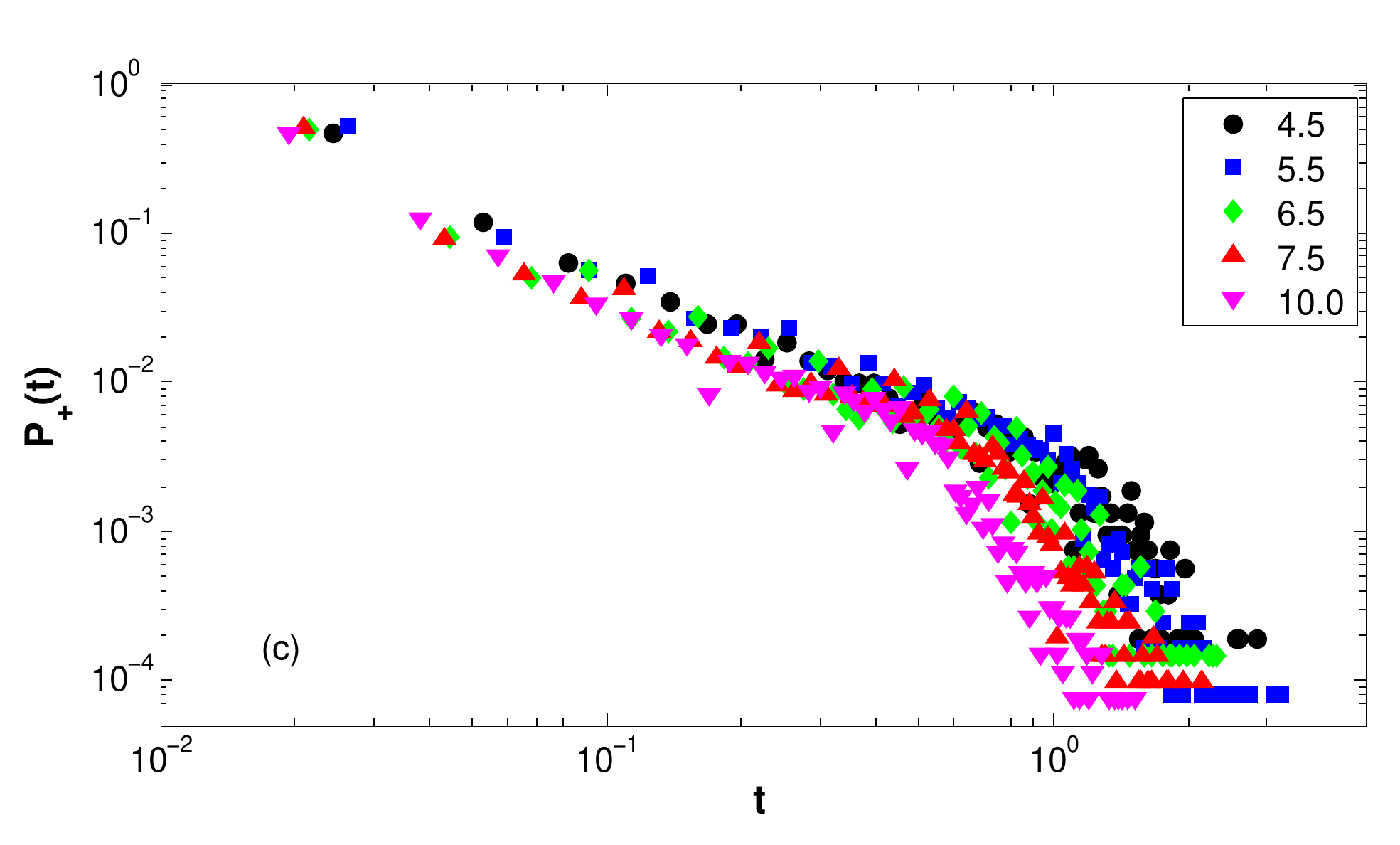}
\includegraphics[scale=0.39]{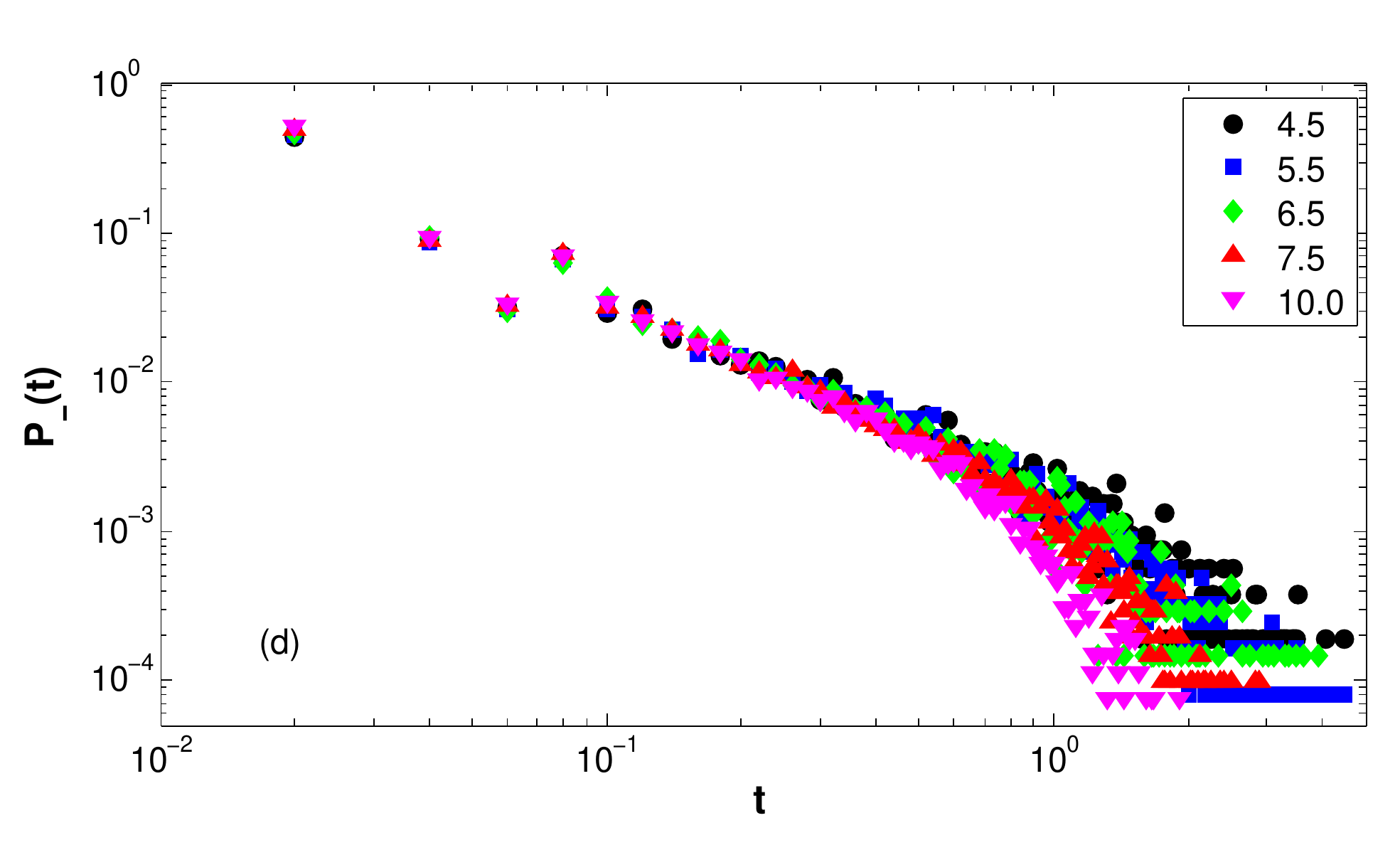}
\caption{\label{fig:firstpassage} (Color Online) Distributions of first passage times, $P_+ (t)$ and $P_-(t)$:  Outlet Box (top panel: a, b)  and Boundary box (bottom panel: c, d). The cutoff times obtained from fits to the tails of these distributions are shown in Table~\ref{firstpassagetimes}.}
\end{figure}

To study the first passage probabilities, we introduce a clipped variable at each box, defined as $s (t) =  \tilde{v}/|\tilde{v}|$, which changes sign at the zeroes of a scaled velocity variable $\tilde{v}$.  This scaled velocity is defined for each box as $\tilde{v} =( v_y - \langle v_y \rangle) / \sigma$  where $\langle v_y \rangle = \mathcal{V}_y (\bm{r} = x_{box},y_{box})$ and $\sigma$ is the standard deviation obtained from the time series, $v_y ((\bm{r}=x_{box},y_{box}),t)$. We then define $P_+ (t)$ ($P_-(t)$)  as the probability that this function does not change sign during a time interval of length $t$, i.e. retains a value $s(t) = +1$ ($s(t) = -1$) and contains no zeroes. In other words, $P_+ (t)$ ($P_-(t)$) measure the probability that a fluctuation that is faster (slower) than the mean, lasts for precisely a time interval $t$.  We examine these processes in the three representative boxes, highlighted in Figure \ref{fig:vfield}, which represent the different regimes of $\mathcal{V}_y ({\bf{r}})$, and where we also look at the distributions of the velocity and its autocorrelation function. These distributions, shown in  Figure~\ref{fig:firstpassage}, appear as power laws cutoff by exponentials for all flow rates, and in all the boxes studied.  The cutoff times increase as the flow rate decreases, implying that the average first passage time increases as the flow slows down, whether we are looking at the time characterizing the switch from above to below the mean ($\langle \tau_+ \rangle$), or or from below to above the mean ($\langle \tau_- \rangle$).  The times $\langle \tau_+ \rangle$ and $\langle \tau_- \rangle$ obtained from exponential fits to the tails of the data are shown  in Table~\ref{firstpassagetimes}.  The distributions unambiguously show an increasingly intermittent flow.  We speculate that the trend in average first passage times indicates some type of localization in trajectory space, akin to proposed mechanisms for the glass transition\cite{garrahan_chandler}.  
 
\begin{table}
\begin{tabular}{| c || c | c | c |c | c | c |}
\hline
&
 \multicolumn{2}{|c|}{ Outlet \ \ } &
\multicolumn{2}{c|}{Boundary} &
\multicolumn{2}{c|}{ Bulk}  \\
\hline
Outlet size & $\tau_+$ & $\tau_-$ & $\tau_+$ & $\tau_-$ & $\tau_+$ & $\tau_-$ \\
\hline
4.5 & 0.382 & 0.442 & 0.365 & 0.707 & 0.435 & 0.427 \\
\hline
5.5 & 0.341 & 0.400 & 0.371 & 0.561 & 0.372 & 0.427 \\
\hline
6.5 & 0.293 & 0.316 & 0.319 & 0.572 & 0.376 & 0.422 \\
\hline
7.5 & 0.246 & 0.274 & 0.261 & 0.393 & 0.375 & 0.400 \\
\hline
10.0 & 0.214 & 0.264 & 0.171 & 0.278 & 0.352 & 0.386 \\
\hline
\end{tabular}
\caption{This table shows, the average first passage times for the velocity at a particular location to go from above to below the mean ($\langle \tau_+ \rangle$), or or from below to above the mean ($\langle \tau_- \rangle$). The rows correspond to different flow rates, and each pair of columns corresponds to the average times at the outlet, boundary, and bulk boxes.}
\label{firstpassagetimes}
\end{table}



\subsection{Velocity PDFs}

\begin{figure}
\includegraphics[width=0.47\textwidth]{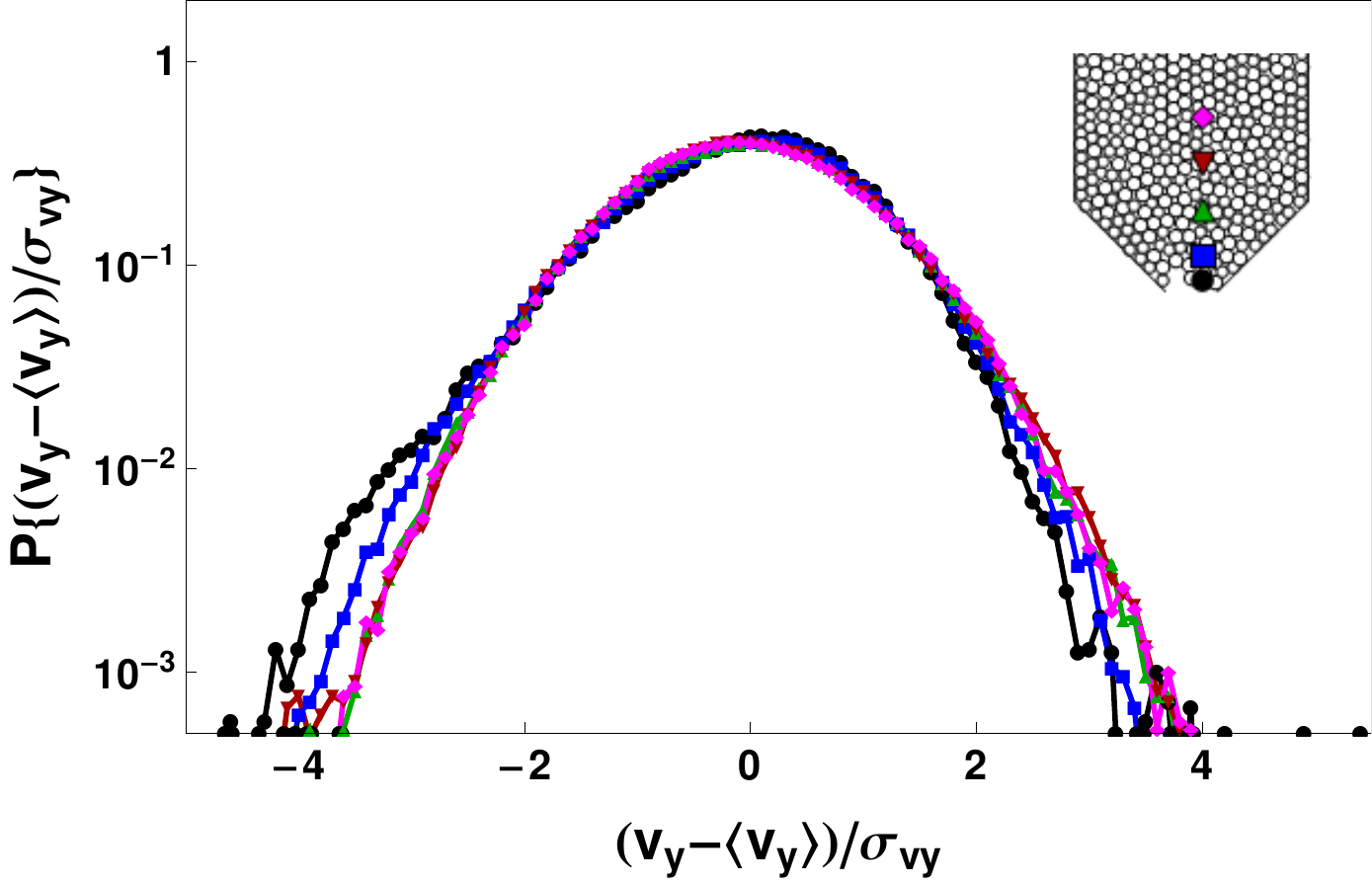}
\includegraphics[width=0.47\textwidth]{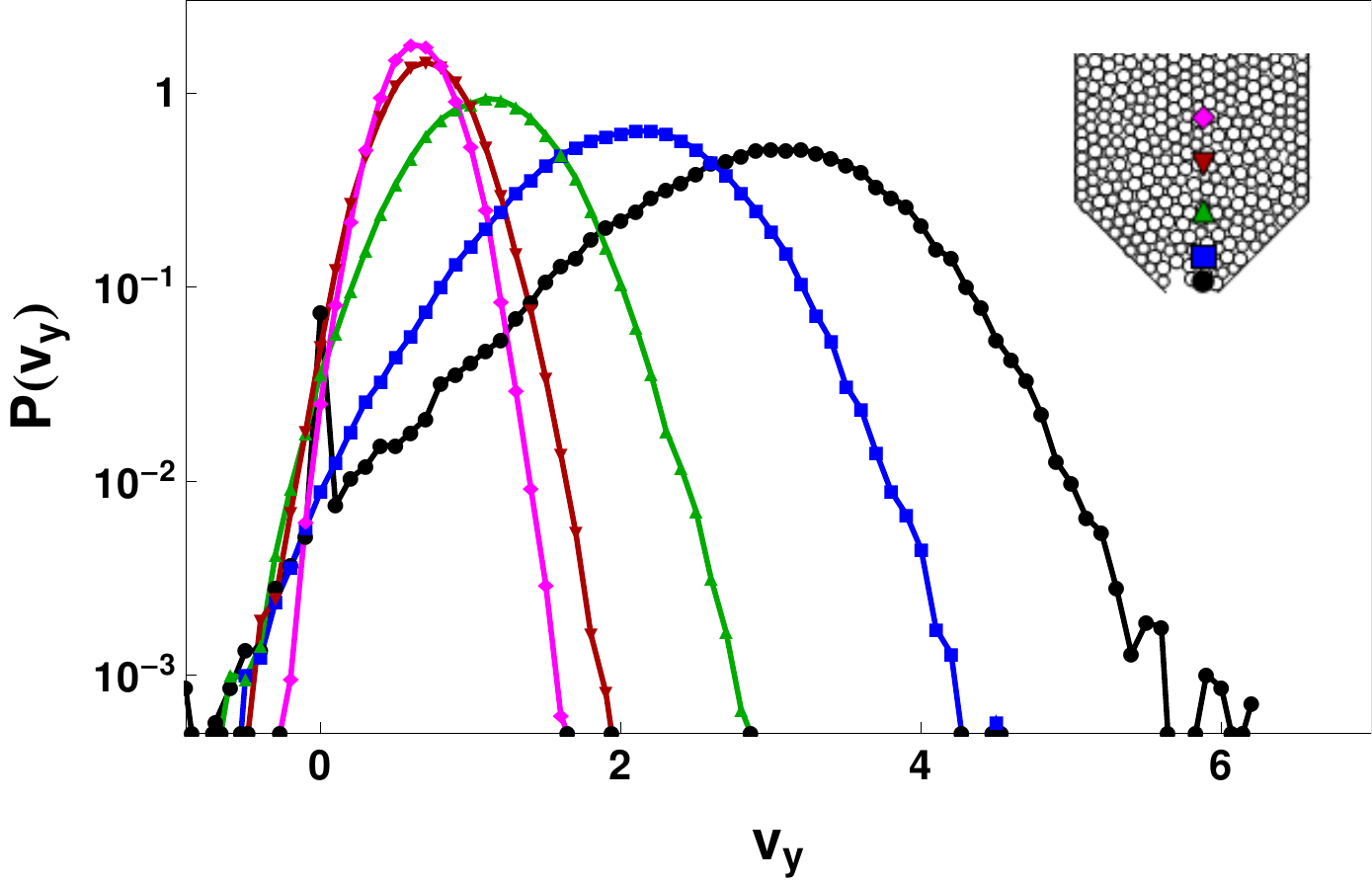}
\includegraphics[width=0.47\textwidth]{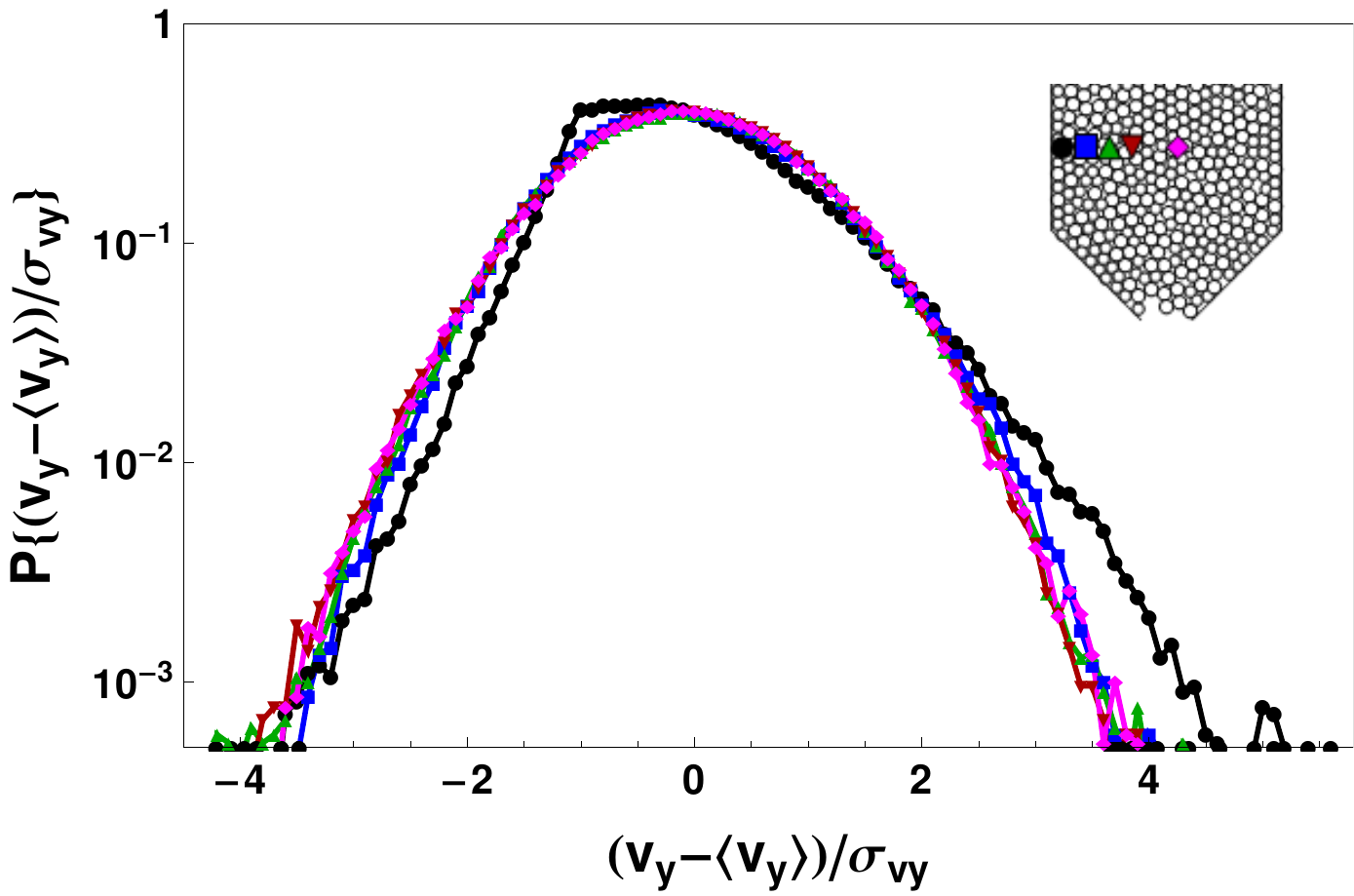}
\includegraphics[width=0.47\textwidth]{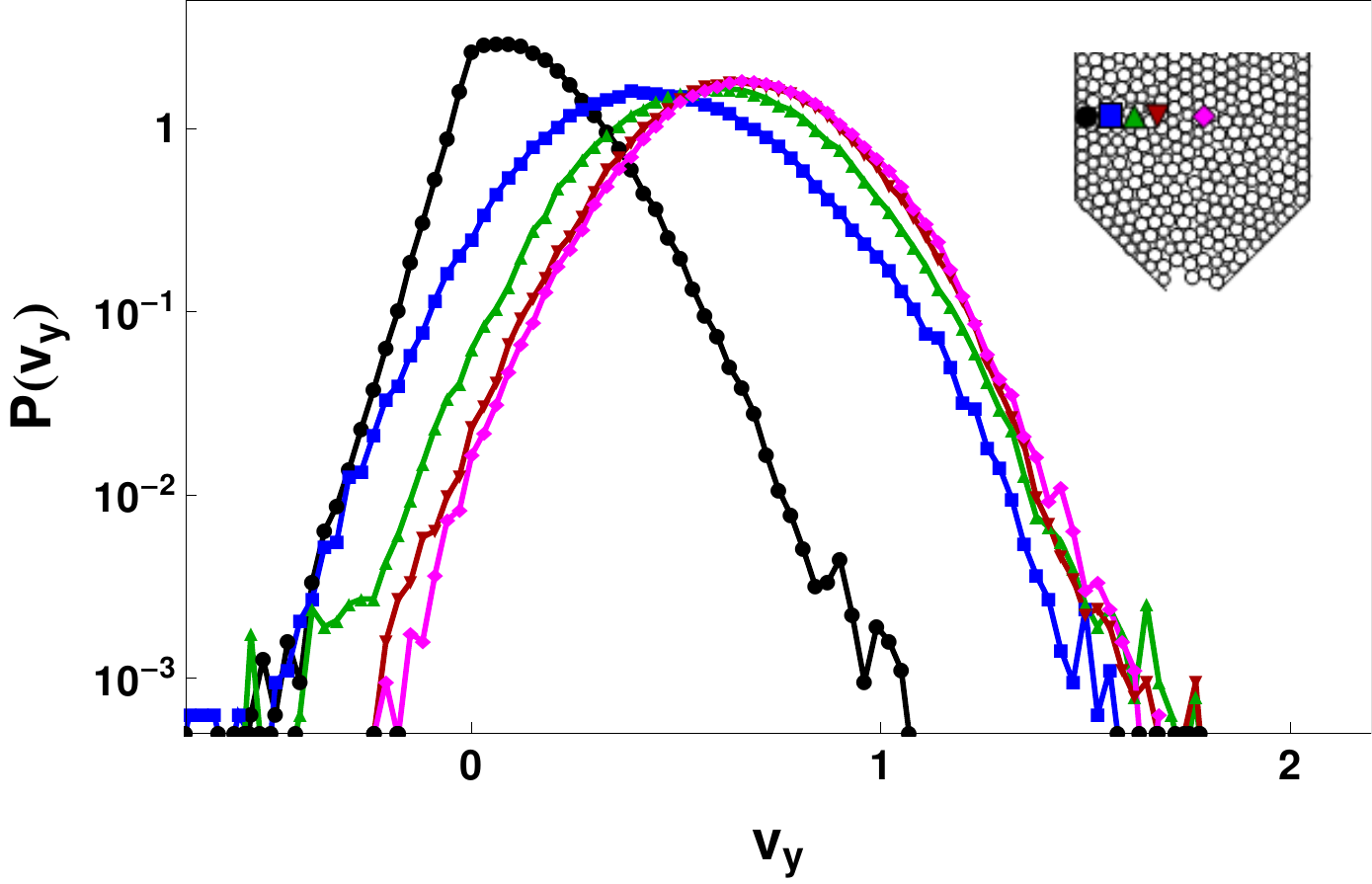}
\caption{\label{fig:vdist} (Color Online) Velocity Distributions at the smallest outlet size, $4.5 d$.  Top Panel:  (Left) The distributions of the scaled vertical velocity $\tilde{v}=( v_y - \langle v_y \rangle) / \sigma$ for the column of boxes shown in the inset; (Right)  distributions of the unscaled velocity $v_y$ for the same set of boxes.  Bottom Panel: (Left) distributions of  $\tilde{v}$ for the row of  boxes shown in the inset; (Right) distributions of the unscaled $v_y$ for the same set of boxes. The corresponding values of the mean, standard deviation and skewness are summarized in Table \ref{mean_deviation}.}
\end{figure}

The probability distributions functions (PDFs) of the vertical components of the velocity become increasingly non-Gaussian as the flow reaches the pre-jamming regime.  A regime of non-Gaussian fluctuations has been observed in experiments on two-dimensional silos below a certain orifice size\cite{janda}. In this section, we analyze the spatial variation of these PDFs.
We have constructed box-specific distributions of $v_y (\bm{r},t)$ and of the scaled velocity variable $\tilde{v}(\bm{r},t) =( v_y(\bm{r},t) - \langle v_y \rangle) / \sigma$ by monitoring the time evolution of the velocity in each box.  Figure \ref{fig:vdist} shows distributions of the scaled and unscaled velocities for a column of boxes parallel to the flow and including the orifice, and a row of boxes transverse to the flow and including the shear layer.   All of the distributions exhibit significant non-Gaussianity, measured quantitatively by the skewness and kurtosis. Table \ref{mean_deviation} summarizes the values of the mean, the standard deviation and skewness for these boxes.  

Different classes of velocity distributions in different regions have been noted in event-driven simulations of flow in three-dimensional silos that are not tapered\cite{drozd_pre08}.   We find marked spatial variation of the velocity fluctuations near the orifice and the vertical walls, whereas the three-dimensional flows showed significant variations only near the top of the hopper.  It is not clear whether these differences stem from dimensionality, the geometry of the hopper or both.  Our results are relevant for experiments performed in a 2D tapered geometry\cite{tang_behringer,gardel}. 

\begin{figure}
\includegraphics[width=0.46\textwidth]{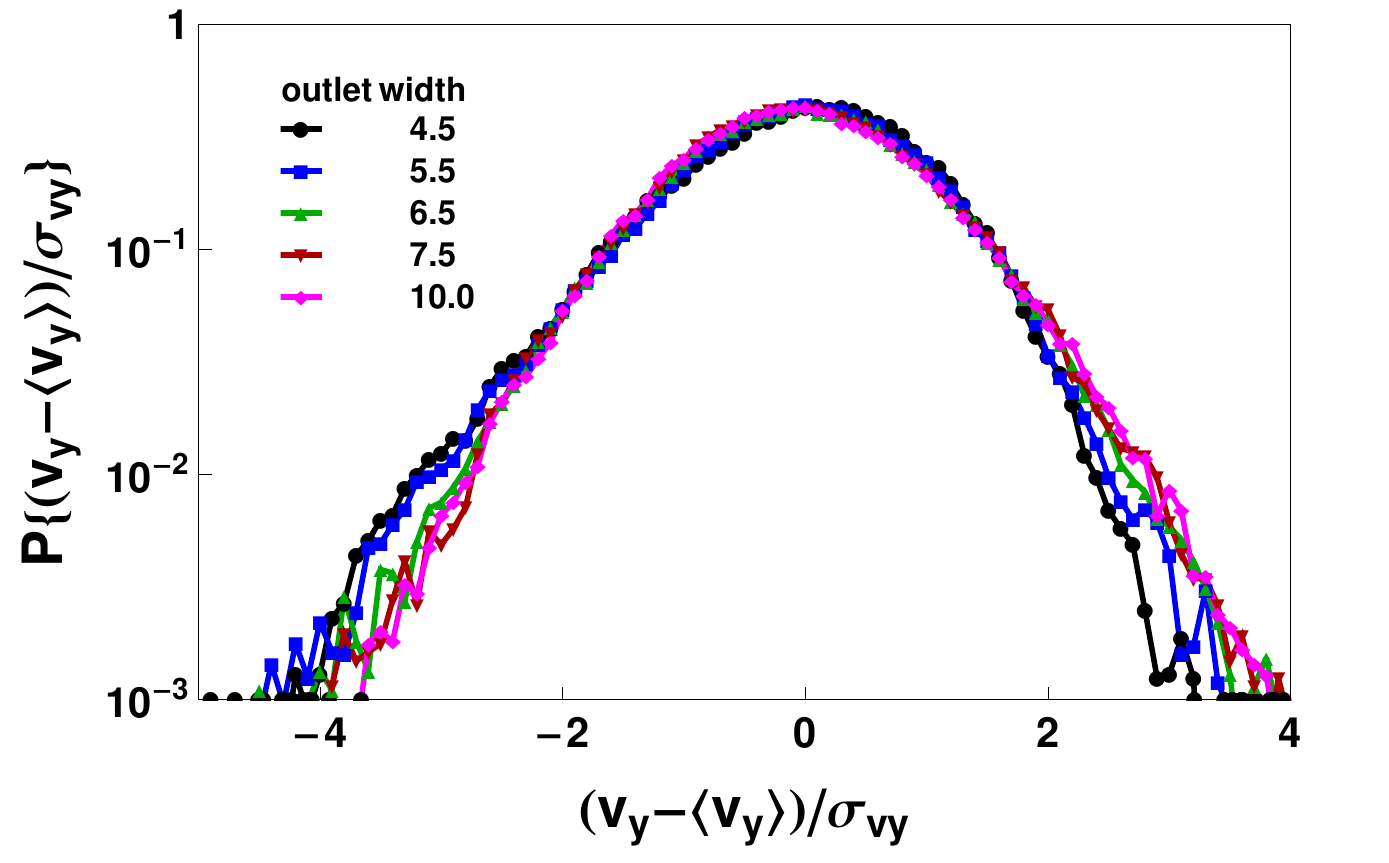}
\includegraphics[width=0.46\textwidth]{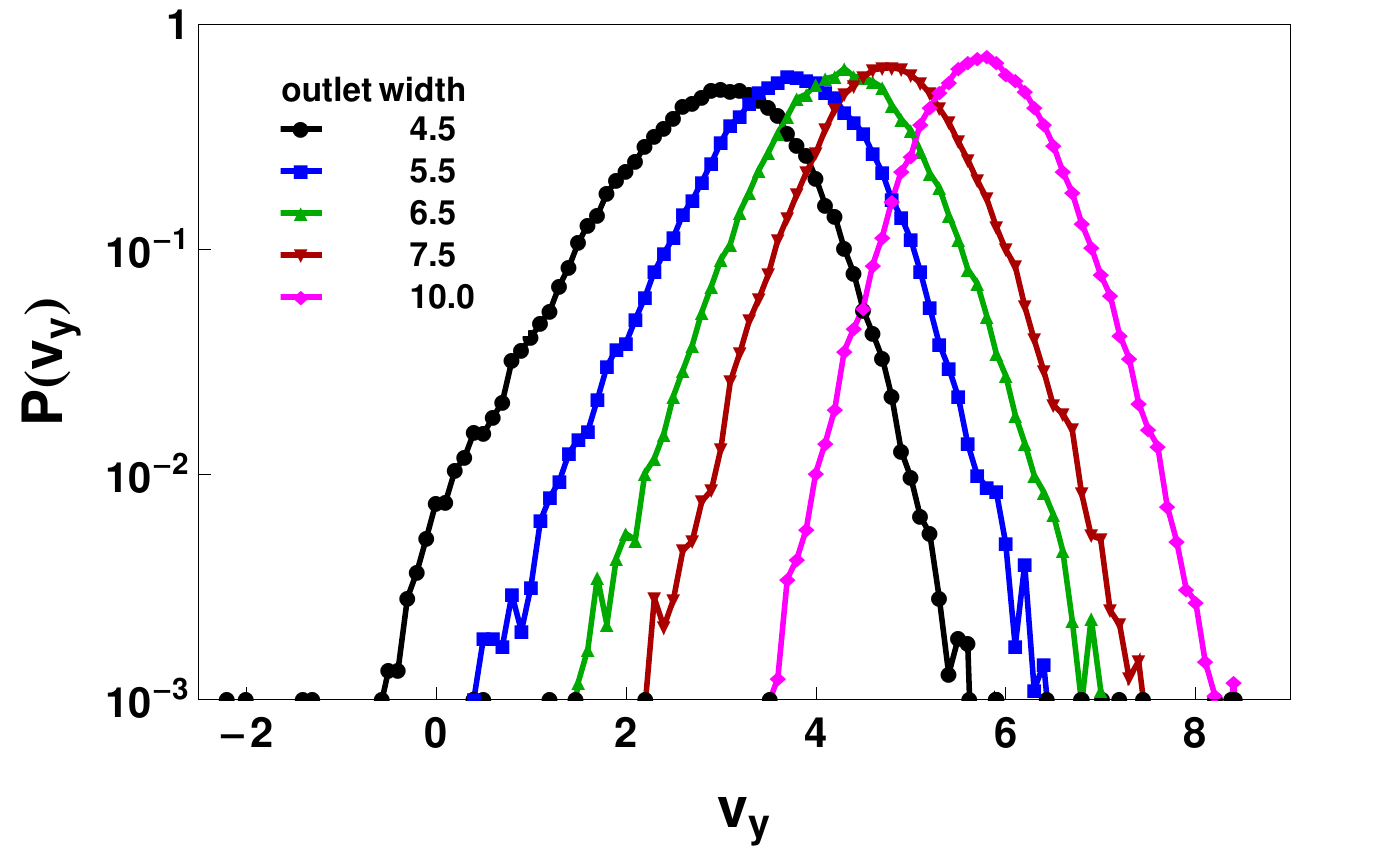}
\includegraphics[width=0.44\textwidth]{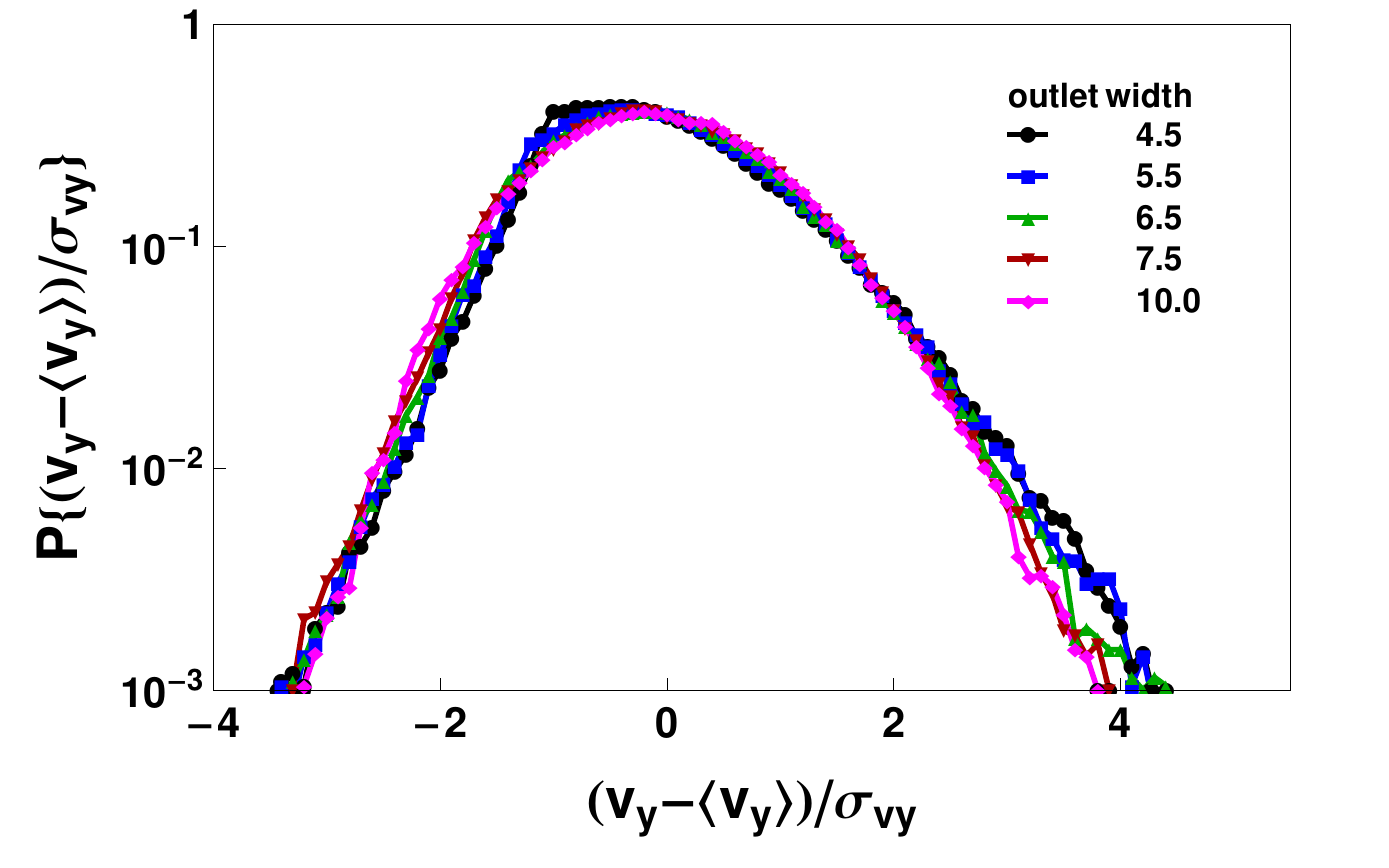}
\includegraphics[width=0.44\textwidth]{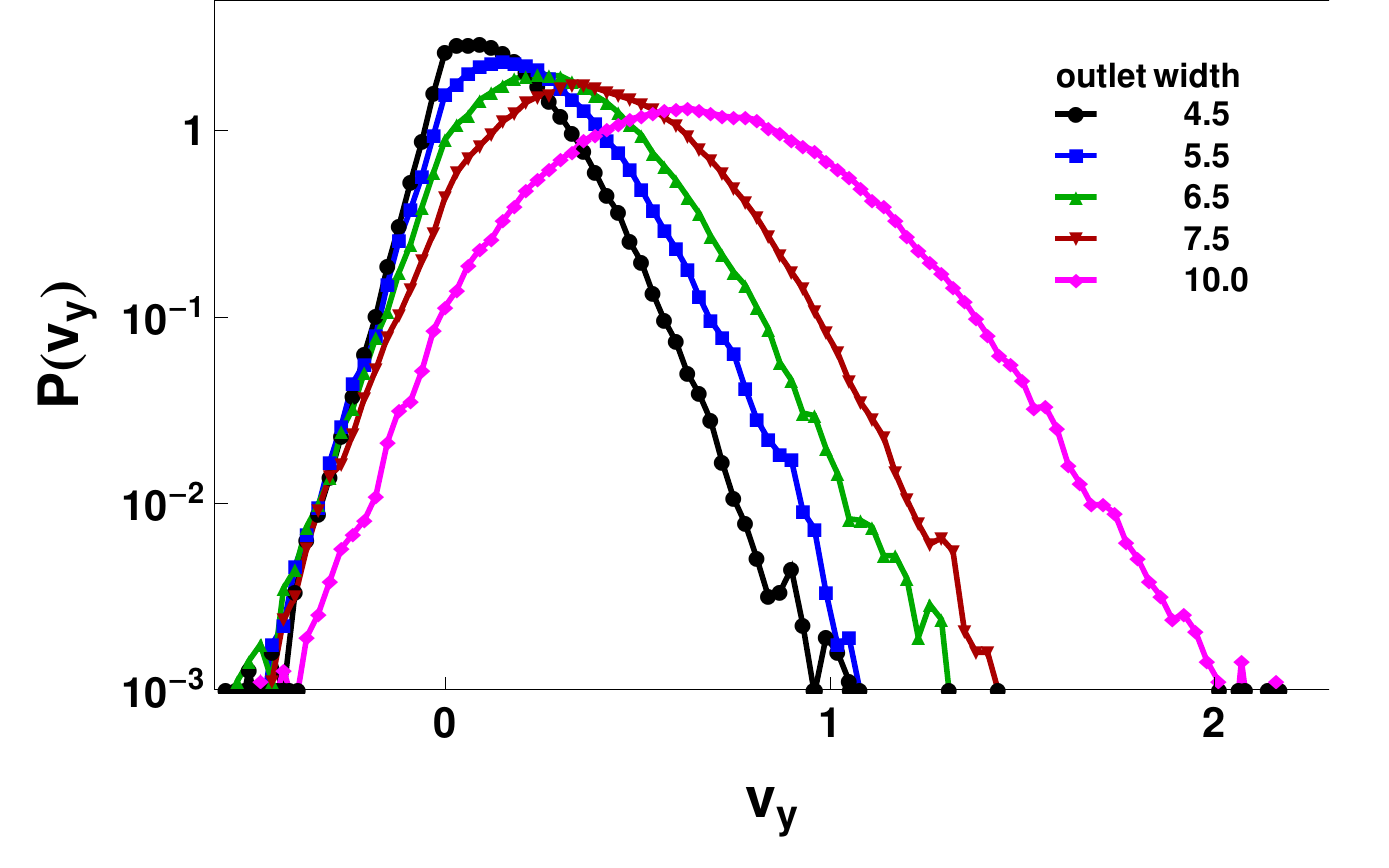}
\caption{\label{fig:vdistflow} (Color Online) Comparison of the velocity distributions  at different flow rates, $v_{flow}$.  Top:  Distributions of the scaled velocity, $\tilde{v}$ (left), and the unscaled velocity, $v_y$ measured at the outlet box.  Bottom:  Distributions of the scaled velocity, $\tilde{v}$ (left), and the unscaled velocity, $v_y$ measured at the vertical boundary box.}
\end{figure}

Figures \ref{fig:vdist} and \ref{fig:vdistflow}, which show velocity distributions at different flow rates, illustrate some remarkable features of the velocity distribution.  At all flow rates, and in all regions that we have examined, except the shear layer spanning $3-4$ grain diameters bordering the vertical wall, the distributions of the vertical component of the velocity are described remarkably well by a universal form, the generalized Gumbel (GG) distribution\cite{gumbel}. The GG is a generalization of distributions found in extreme value statistics\cite{gumbel} that appears to be more broadly applicable: it has been observed in 
a simulated low density granular gas of hard disks\cite{brey,mounier} where it describes the energy distribution, and in sheared systems close to jamming where it describes entropy fluctuations\cite{sood}. The GG distribution is entirely determined by moments of the measured distribution, namely the mean, skewness and standard deviation:
\begin{equation}
\Pi(\tilde{v}) = K e^{a[x-e^x]}, \  \   \text{with  } x = -b(\tilde{v} + z) ~.
\label{eqn:gumbel}
\end{equation}
Here $K, ~b, ~z$ are all functions of the parameter $a$, which is in turn related to the skewness of the distribution: $|\langle \tilde{v}^3 \rangle| = 1/\sqrt{a}$. The functions appearing in Equation \ref{eqn:gumbel} depend only on $a$:
\begin{equation}
b(a) = \sqrt{d^2 \ln\Gamma (a) \over da^2}, \   \  z(a) = {1 \over b} \left( \ln (a) - {d \ln \Gamma (a) \over da} \right) , \  \  K(a) ={ b a^a \over \Gamma (a)} 
\end{equation}
where $\Gamma (a)$ is the Gamma function.   

It should be emphasized that the GG is not a fitting form, but is simulated from measured properties, and compared to the measured distribution. As an example, the figure in the right panel of Table~\ref{mean_deviation} shows a comparison between the best Gaussian fit (blue, dashed), the GG (red, solid), and the measured distributions of $\tilde{v}$ at the outlet.  At a given flow rate, the parameter $a$ characterizing the GG, is smallest closest to the boundaries and increases towards the center of the hopper.  The distributions are, therefore closer to a Gaussian at the center where the flow rate is highest. The skewness also changes sign as a function of vertical position of the box, being negative and large at the orifice and positive but smaller in the central region.   A negative skewness indicates a larger weight in the distribution below the peak than above it.   At a given position in the hopper, the GG parameter decreases with decreasing outlet width. The GG form describing the measured velocity distributions are, therefore, farthest from a Gaussian at the slowest flow rate and closest to the orifice.  The GG parameter $a$ provides a quantitative measure of the variation of the fluctuation-statistics with flow rate.
\begin{table}
\parbox{.28\linewidth}{
\begin{tabular}{ | c || c | c | c |}
\hline
 & $\langle v_y \rangle$ & $\sigma$ & $\langle\tilde{v}^3\rangle$\\
\hline\hline
{\tiny \bf{ Bottom}} & 2.935 & 0.864 & -0.594 \\
\hline
  & 2.076 & 0.638 & -0.196 \\
\hline
{\tiny \bf{ to}}  & 1.157 & 0.428 & 0.0064 \\
\hline
 & 0.763 & 0.278& 0.054 \\
\hline
{\tiny \bf{ Top}} & 0.700 & 0.222 &0.071 \\
\hline
\end{tabular}}
\parbox{.28\linewidth}{
\begin{tabular}{ | c || c | c | c |}
\hline
 & $\langle v_y \rangle$ & $\sigma$ & $\langle\tilde{v}^3\rangle$\\
\hline\hline
{\tiny \bf{ Left}} &0.156 & 0.148 & 0.686 \\
\hline
  & 0.479 & 0.253 & 0.189 \\
\hline
{\tiny \bf{ to}} & 0.615 & 0.239 & 0.068 \\
\hline
 & 0.681 & 0.224 & 0.051 \\
\hline
{\tiny \bf{ Center}} & 0.700 & 0.222 & 0.071 \\
\hline
\end{tabular}
}
\parbox{0.4\textwidth}{
\begin{tabular}{|c|}
\hline
\includegraphics[width=0.4\textwidth]{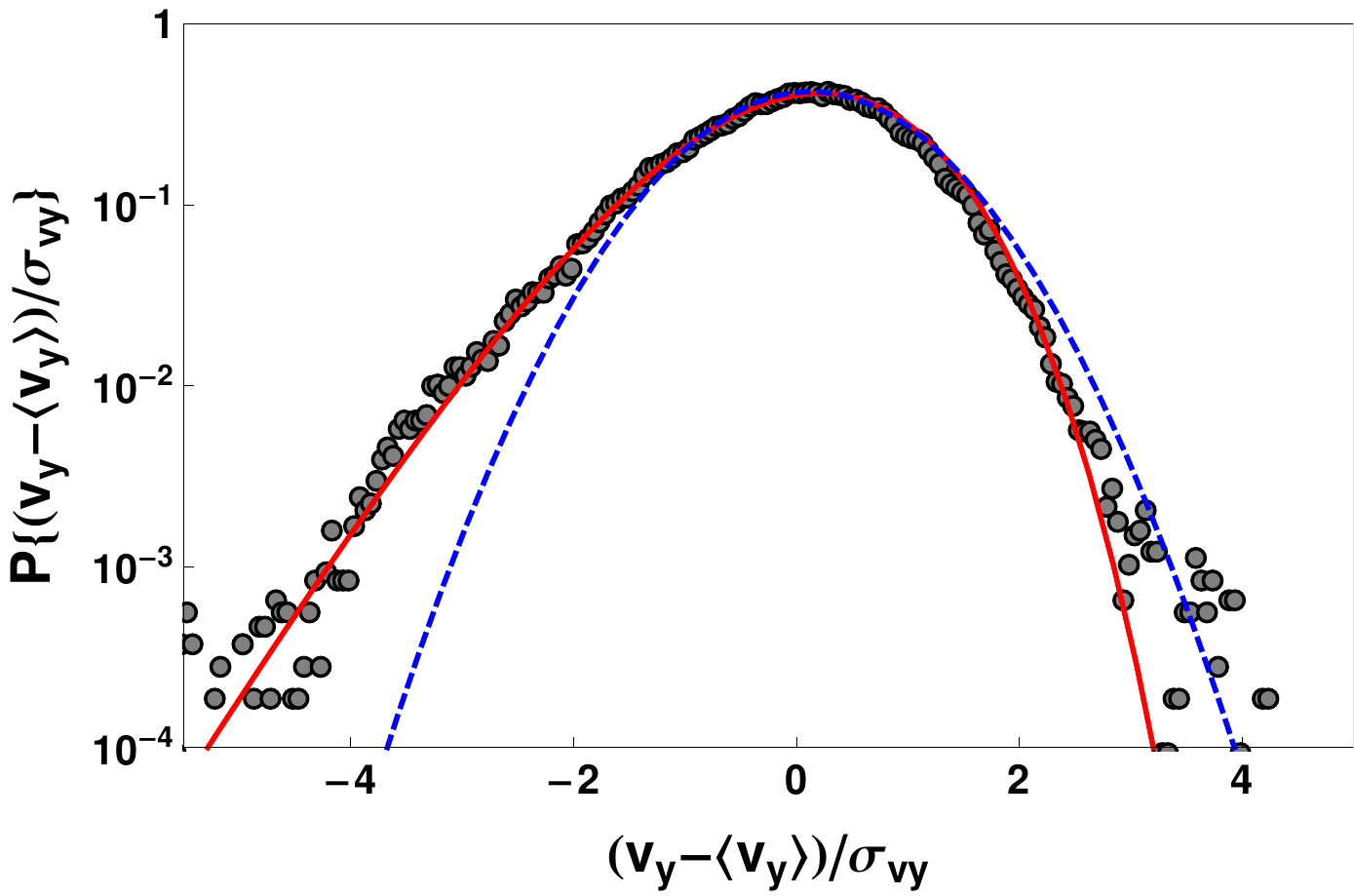}\\
\hline
\end{tabular}
}
\caption{The mean,  $\langle v_y \rangle$, standard deviation, $\sigma = \sqrt{\langle \tilde{v}^2 \rangle}$,  and skewness, $\langle \tilde{v}^3\rangle$ of the velocity distributions shown in Figure \ref{fig:vdist}.   The panel on the right compares the distribution of $\tilde{v}$ at the orifice to the best Gaussian fit (blue, dashed) and the simulated GG distribution (red, solid).} 
\label{mean_deviation}
\end{table}

The appearance of a universal, non-Gaussian distribution suggests that there is an underlying physical principle leading to this particular form of non-Gaussian behavior.  We do not yet have any insight into this physical mechanism, however, we expect that comparing to other near-jammed systems will  be a productive avenue to pursue. 

The shape of the distribution within the shear layer at the vertical wall, while strikingly non-Gaussian, cannot be described by the GG form either. There are significant negative velocity events that signal particles moving opposite to the direction of the driving, and the distribution of these negative events is very close to exponential. The edge of this exponential behavior seems to be pinned at zero velocity for a range of outlet openings, as seen in the lower panel of Figure~\ref{fig:vdistflow}. 
The appearance of significant negative-velocity events seems to be a hallmark of near-jammed systems\cite{sood,nitin}. Intriguingly, there is remarkable similarity between the velocity distribution in the shear layer of our model hopper flow and that of a polar granular rod moving in a vibrated bed of granular spheres\cite{nitin}.    Negative events are also observed at the orifice of the hopper, however, their contribution to the velocity distribution appears to be consistent with the shape of the GG, as shown in the right panel of Table~\ref{mean_deviation}. Zero flow events at the outlet box, not shown in Table~\ref{mean_deviation}, appear as a narrow peak superposed on top of the GG distribution as seen earlier in Figure~\ref{fig:timetrace}; these occur when there are no particles in the outlet box in a given time interval.

In the next section, we describe our analysis of a jam that persists for $10^7$ collisions, and provide evidence for the appearance of spatial structures with growing length scales, including vortices.

\subsection{Jam Event} So far, our discussion has focused on fast flows and the pre-jamming flow which is an interesting NESS that cannot be cleanly labeled as a fluid or a solid.  In this section, we analyze a jam where the the flow stops for a time interval spanning $10^7$ collisions. This event offered us the rare opportunity of studying a NESS that is as close to a solid as possible in purely collisional flows.  
 
\begin{figure}
\includegraphics[scale=0.25]{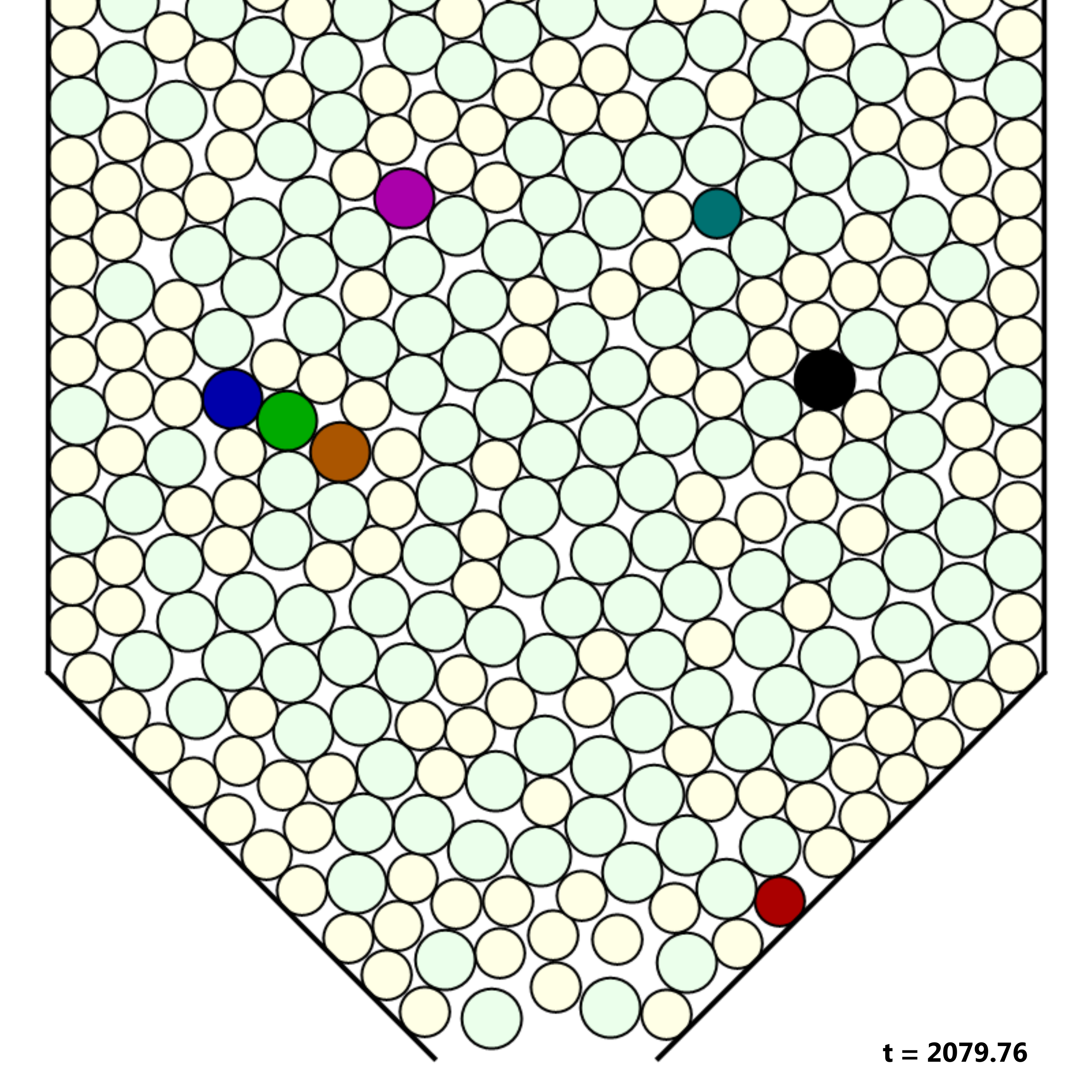}
\includegraphics[scale=0.25]{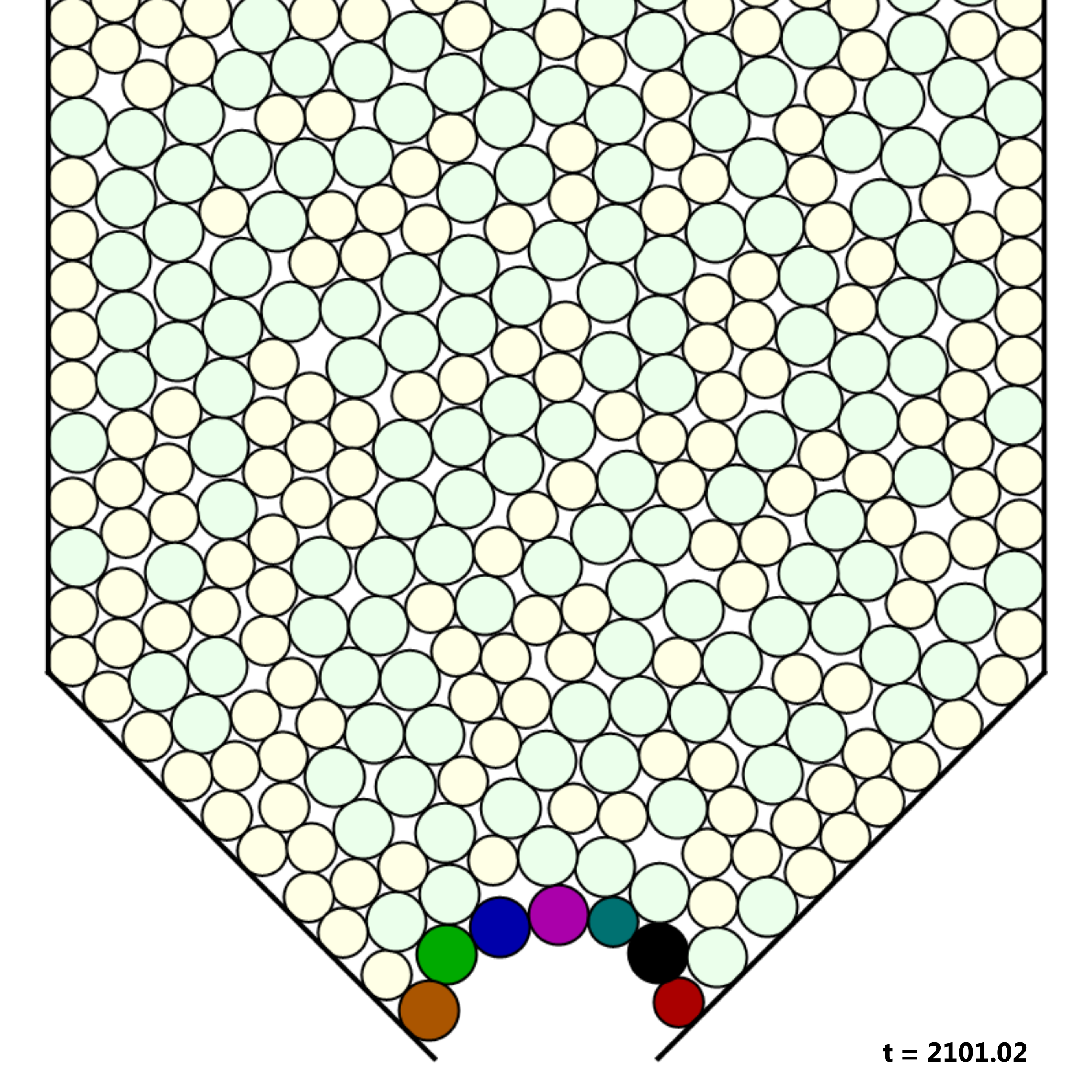}
\caption{\label{fig:jam} (Left) (Color Online) A snapshot taken approximately $10^3$ collisions before the arch formation, showing the position of the particles that end up forming the arch (Right) responsible for the jam. Tracking the motion of these particles, there motion is found to bear a striking resemblance to the process of arch formation seen in experiments\cite{tang_behringer}.}
\end{figure}

We observed that the jam was due to the formation of a classic arch at the outlet.  We tracked the particles that ended up in the arch, Figure~\ref{fig:jam}, and found that arches are formed at the outlet and not pre-formed upstream in the flow. In the simulation, as in experiments on photoelastic beads\cite{tang_behringer}, arch-forming particles originate in disparate regions of the system. These similarities with experiments suggest that a jam in a collisional flow might not be that different from one in a physical system with extended contacts, and encouraged us to analyze the anatomy of this jam.
We studied two aspects of the jam: (1) appearance of vortices in the velocity field and (2) geometry of spatio-temporal clusters.

\begin{figure}
\hspace{0.4in}
\includegraphics[width = 0.4\textwidth]{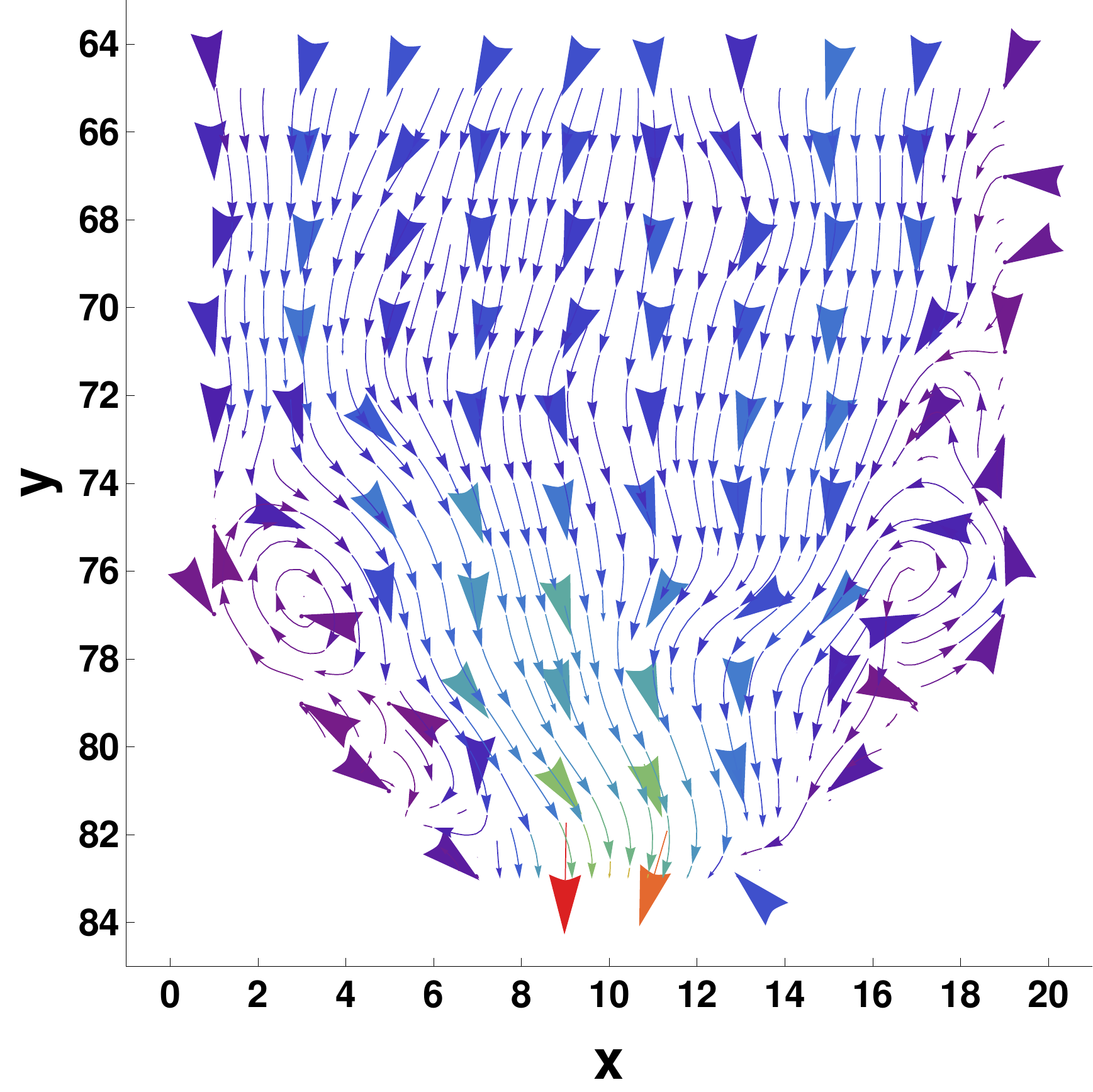}
\includegraphics[width = 0.52\textwidth]{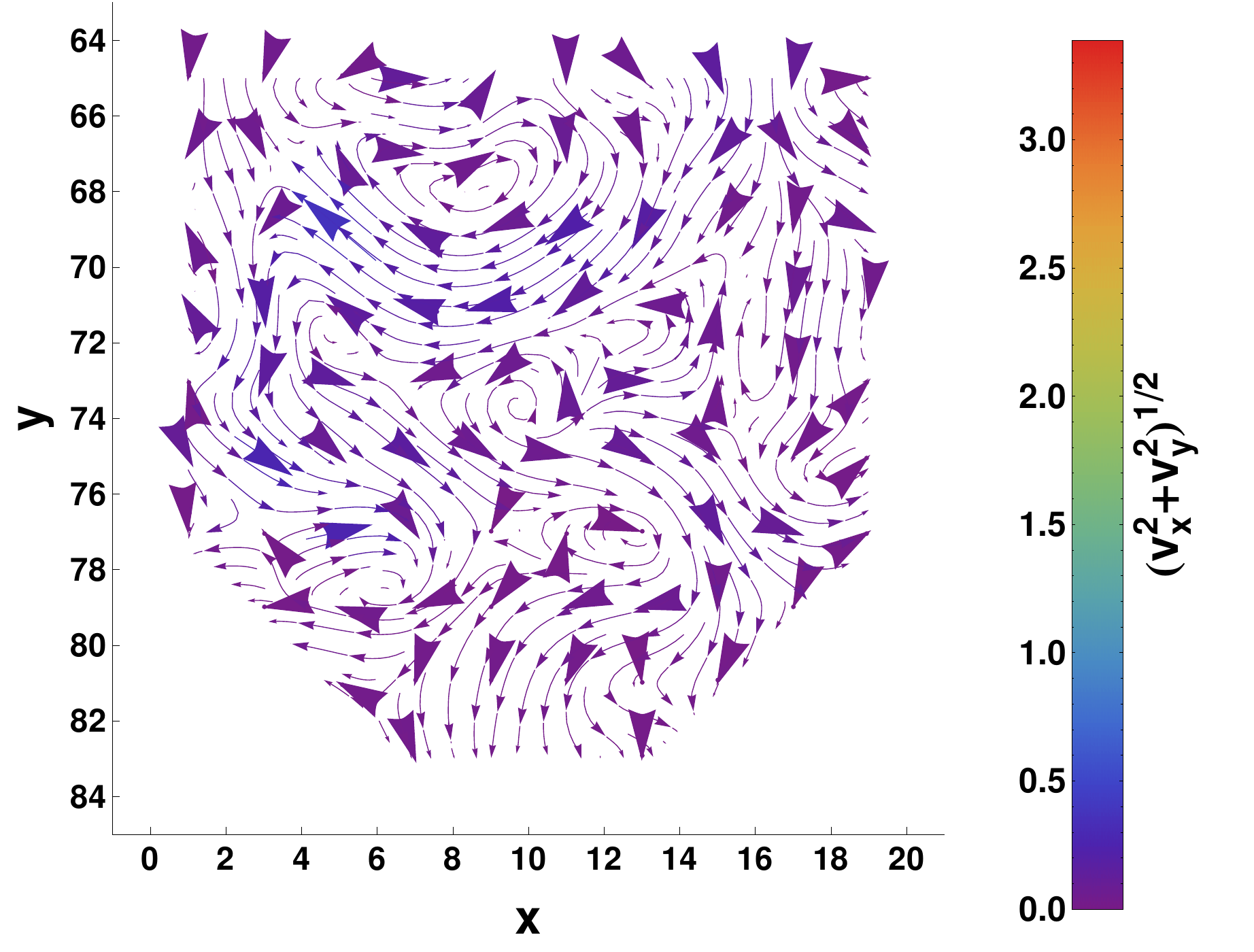}
\includegraphics[width = 0.43\textwidth]{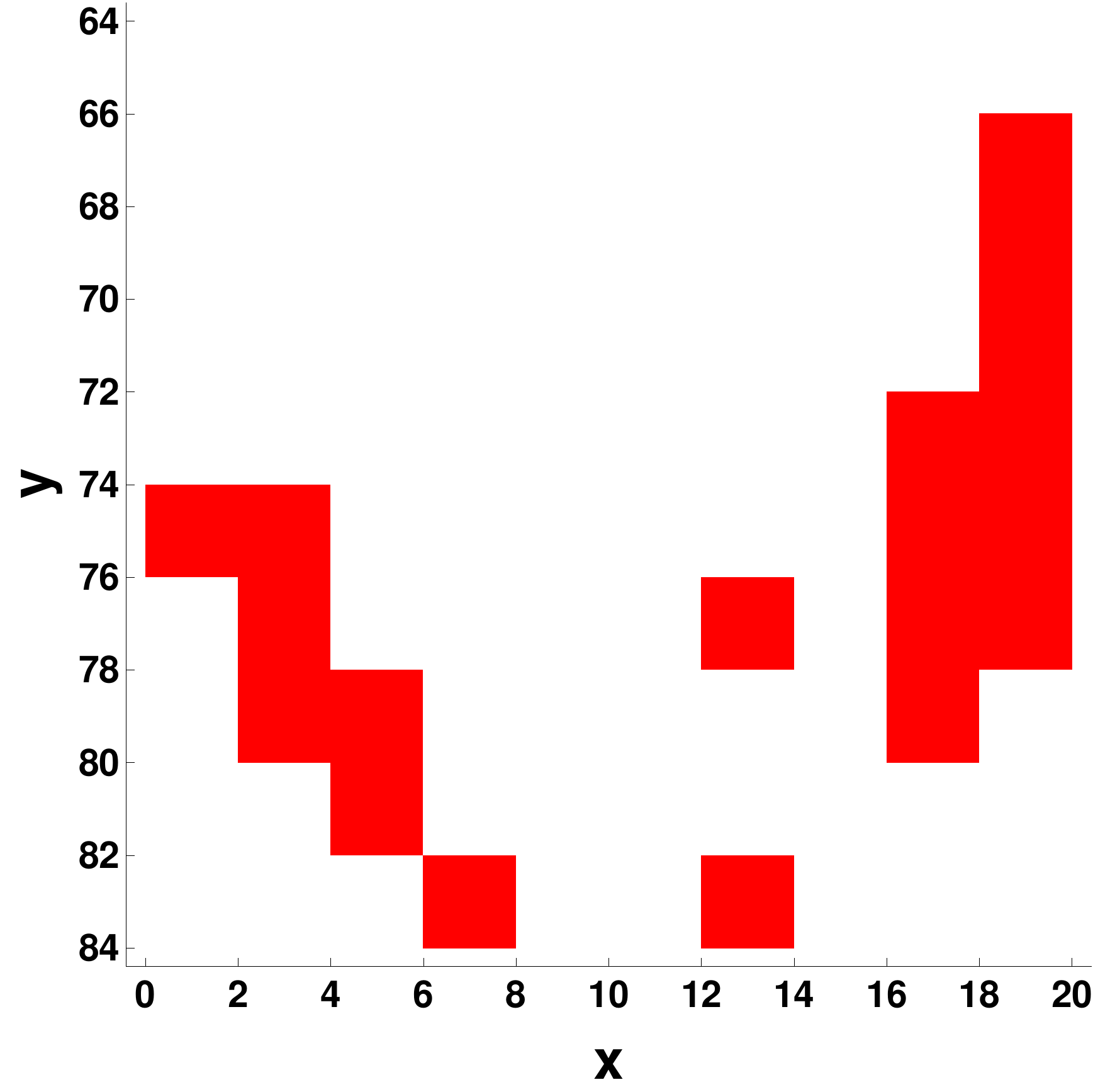}
\includegraphics[width = 0.43\textwidth]{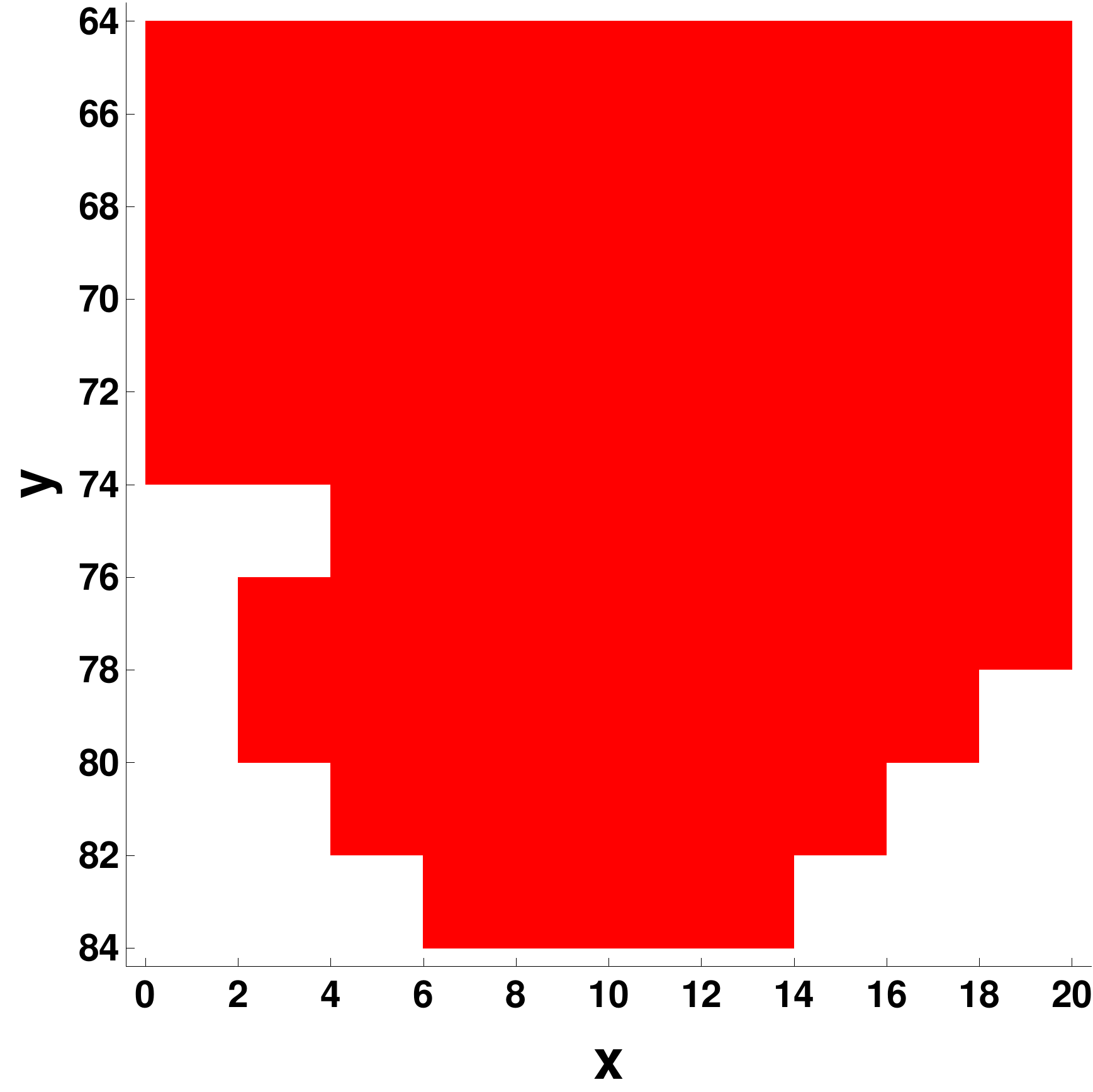}
\caption{\label{fig:vorticity} (Color Online) The upper panel of this figure shows the instantaneous velocity field in the lower region of the hopper at two times leading up to the jam. The lower panel corresponds to the same two instants in time, and highlights boxes in which the instantaneous vertical velocity is less than half the mean in that box. The figures on the right are closer in time to the jam.}
\end{figure}

We tracked the time evolution of the velocity field ${\bm v}$ through the formation and dissolution of the jam.  The upper panel of Figure~\ref{fig:vorticity} shows the velocity field at two instants in time leading up to the jam, and clearly demonstrates the increasing vorticity of the flow. Comparing the snapshot on the left to the one on the right, which is at a later time, we see that vortices nucleate at the edges of the hopper, and propagate into the bulk. We are currently in the process of measuring this behavior systematically\cite{md_inprep}. To our knowledge, vorticity of self-organized flows such as the hopper flows have not been investigated much nor have they been correlated with jamming.  In contrast, length scales associated with vorticity have been studied in chute flows\cite{ertas_halsey}, and have been related to jamming and flow in that geometry through a scaling analysis.  Our results suggest that vorticity could be related to the dynamical arrest of hopper flows.  The negative velocity events that we observed in the intermittent flow regime can also be related to vortices, but are transient,  and unlike the ones in Figure~\ref{fig:vorticity}, do not have any significant spatial extent.

\begin{figure}
\includegraphics[width=\textwidth]{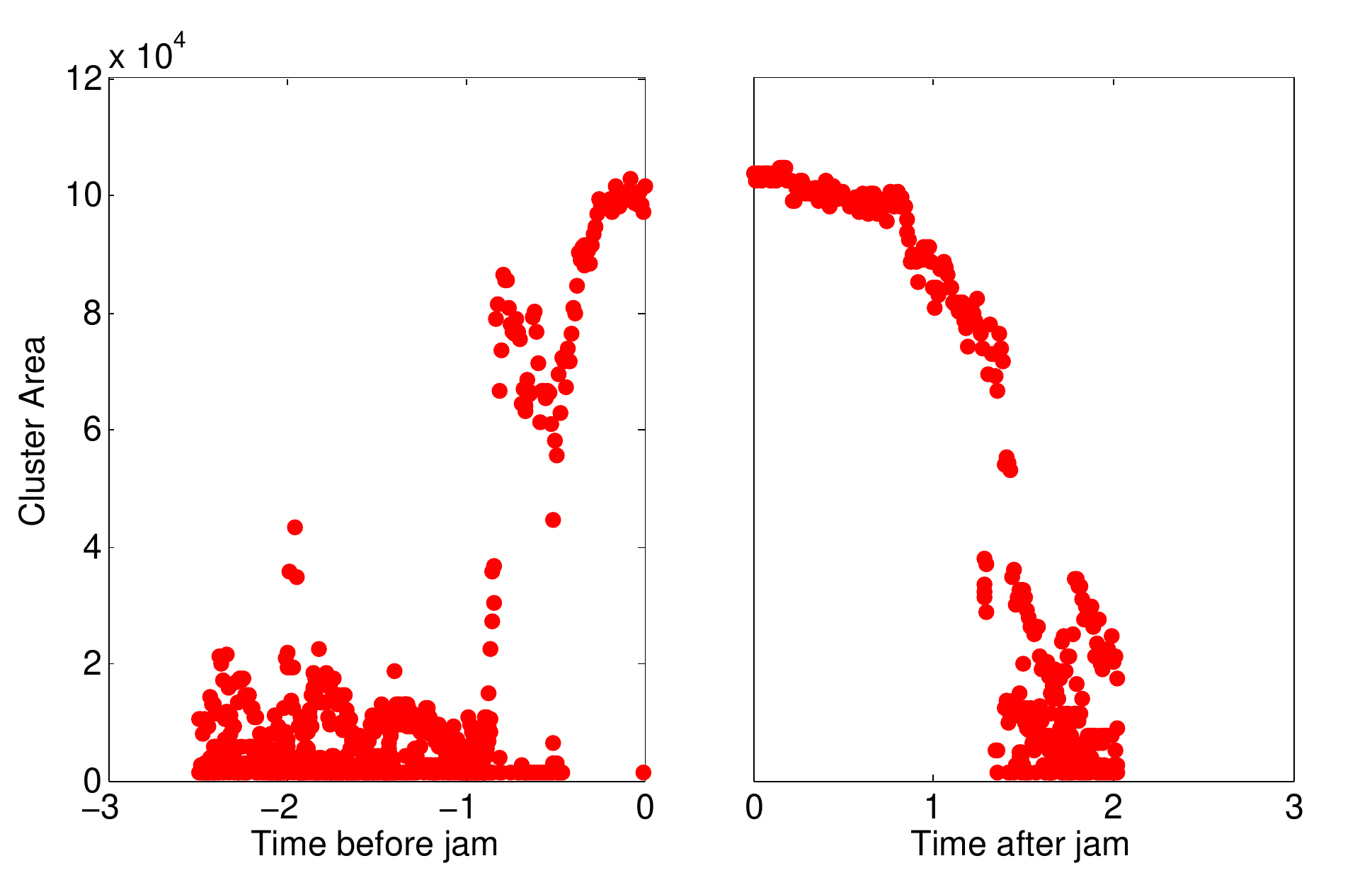}
\caption{\label{fig:jamarea} Scatter plot of the area covered by connected clusters, defined by boxes in which the flow is less than half the average, plotted against time before (on the left) and after (on the right) the jamming event. The duration of the jam that separates the two figures is about 24 simulation times.}
\end{figure}

We have also analyzed the spatial extent of `slow clusters'.  We define slow boxes as ones that have, at a given instant, a vertical velocity smaller than half of the average flow rate in the box. Clusters are identified by nearest-neighbor connectivity and then tracked as a function of time.  The lower panel of Figure \ref{fig:vorticity} shows two snapshots of cluster formation in the time leading up to the jam.  Each of these cluster images corresponds to the same instant in time as the vorticity image in the panel immediately above. The two images on the left demonstrate that there is vorticity associated with the regions where we see large clusters. The images on the right are closer to the instant of the jam, and the cluster spans nearly the entire region. To obtain quantitative information, we plot cluster area as a function of time both during the formation and the breakup of the jam. Figure~\ref{fig:jamarea} shows that formation and breakup of the jam is signaled by a rapid change in cluster area.  This is not surprising and corroborates the picture that a stable arch at the outlet supports the weight of the grains above while large regions of slow flow develop.   

The clusters that are observed during the jamming event  are an example of a dynamically generated structure.  Similar structures have been seen in experiments\cite{gardel}, so we do not think this is an artifact of our simulation. In our collisional model, these slow flow regions are nowhere near static, instead, the grains continually undergo collisions.  The clusters retain their identity while their constituents participate in a highly correlated yet rapid dynamics.  Our ultimate aim is to understand the dynamical principles that lead to the emergence of these structures, and their stability. In the next phase of this work, we will investigate the formation and breakup of stable, long-lasting jams by generating many realizations of our simulation at the smallest opening\cite{md_inprep}, and by using molecular dynamics simulations with particles interacting via Hertzian contacts\cite{Silbert_Grest}.

\section{Conclusion} We have presented an analysis of the jamming transition in purely collisional 2D hopper flows.  In our simulations, we identify new dynamical signatures of the approach to jamming in this self-regulated flow.  In particular, we analyze features that are strongly influenced by  the boundaries, and find a striking change in the nature of the flow as the opening size is reduced.    The spatial distribution of velocity-autocorrelation times shows that, in a pre-jamming regime,  the flow has a distinctly heterogeneous character.  There is a glassy layer at the walls, characterized by anomalously slow velocity relaxations, coexisting with a central flowing region. This is in contrast to higher flow rates, where the flow resembles a thermal fluid with strong spatial gradients.  The flow is strongly intermittent in the pre-jamming regime, with well-defined periods of slow and fast flow.  From measurements of first-passage time distributions, we deduce that velocity fluctuations that dip below the  mean persist longer as jamming is approached.  The effect is strongest at the shear layer, where we also see the development of a glassy region.   


We also observe marked vorticity in the flow as an extended jam forms and then breaks up.  The vortices nucleate at the corners of the hopper and extend inwards, ultimately causing complete arrest.  The vorticity is spatially correlated with the solid-like regions, suggesting that the slow relaxation of the velocity is related to the vorticity.   In our earlier work\cite{ally_pre}, we saw clear evidence of stress chains in this purely collisional flow. These stress chains were predominantly anchored on the side walls, and it is plausible that they are being supported by the glassy layer. In future work we will investigate the mechanism by which the more glassy boundary layer sheds vortices into the free-flowing bulk layer to generate jams. One question that arises is whether extended contacts are necessary for the onset of jamming; our results here suggest otherwise.  We will carry out molecular dynamics simulations and compare them to the event-driven results in order to fully answer this question. 

\begin{acknowledgments}
We wish to acknowledge useful discussions with Narayanan Menon and Aparna Baskaran. ST acknowledges the support of a summer research grant from Western New England University. BC and MD acknowledge the support of NSF-DMR 0905880. 
\end{acknowledgments}


\begin{thebibliography}{99}
\bibitem{grain_silo} See, for example, J. Nielson, Phil. Trans. R. Soc. Lond. A \textbf{356}, 2667 (1998).
\bibitem{liunagel} A. J. Liu and S. R. Nagel, Nature \textbf{396}, 21-22 (1998).
\bibitem{to_lai} K. To, P-Y. Lai, and H. K. Pak, Phys. Rev. Lett. \textbf{86}, 71 (2001); K. To and P-Y. Lai, Phys. Rev. E \textbf{66}, 011308 (2002).
\bibitem{garcimartin:2010pre} A. Garcimartin, I. Zuriguel, L.A. Pugnaloni, and A. Janda, Phys. Rev. E \textbf{82}, 031306 (2010).
\bibitem{tang_behringer} J. Tang and R. P. Behringer, Chaos \textbf{21}, 041107 (2011);  J.Tang, S. Sagdiphour, and R.P. Behringer, Powders and Grains 2009: Proceedings of The 6th International Conference on Micromechanics of Granular Media, \textbf{1145}, 515 (2009).
\bibitem{thomas_durian} C. C. Thomas and D. J. Durian, arXiv:1206.7052v1.
\bibitem{zuriguel:2011prl} I. Zuriguel, A. Janda, A. Garcimartin, C. Lozano, R. Arevalo, and D. Maza, Phys. Rev. Lett. \textbf{107}, 278001 (2011).
\bibitem{jnb_rmp} H. M. Jaeger, S. R. Nagel, and R. P. Behringer, Rev. Mod. Phys. \textbf{68}, 1259 (1996).
\bibitem{gardel} Emily Gardel, E. Keene, S. Dragulin, N. Easwar and N. Menon, arXiv cond-mat/0601022; E. Gardel, E. Sitaridou, K. Facto, E. Keene, K. Hattam, N. Easwar and N. Menon, Phil. Trans. R. Soc. A \textbf{367}, 5109 (2009).
\bibitem{longhi}  E. Longhi, N. Easwar and N. Menon, Phys. Rev. Lett. \textbf{89}, 045501 (2002).
\bibitem{ally_epl04} A. Ferguson, B. Fisher, and B. Chakraborty, Europhys. Lett. \textbf{66}, 277 (2004).
\bibitem{ally_pre} A. Ferguson and B. Chakraborty, Phys. Rev. E \textbf{73}, 011303 (2006).
\bibitem{shub_pre} S. Tewari, B. Tithi, A. Ferguson and B. Chakraborty, Phys. Rev. E \textbf{79}, 011303 (2009).
\bibitem{Ally_07} A. Ferguson and B. Chakraborty, {Europhys. Lett.}, \textbf{78}, 28003 (2007).
\bibitem{pouliquen_gutfraind} O. Pouliquen and R. Gutfraind, Phys. Rev. E \textbf{53}, 552 (1996).
\bibitem{losert} W. Losert, L. Bocquet, T. C. Lubensky, and J. P. Gollub, Phys. Rev. Lett. \textbf{85}, 1428 (2000).
\bibitem{denniston_li} C. Denniston, and H. Li, Phys. Rev. E \textbf{59} 3289 (1999).
\bibitem{bennaim_redner} E. Ben-Naim, S.Y. Chen, G. D. Doolen, and S. Redner, Phys. Rev. Lett. \textbf{83}, 4069 (1999).
\bibitem{brito_wyart} C. Brito and M. Wyart, J. Chem. Phys. \textbf{131},024504 (2009).
\bibitem{donev_torquato} A. Donev, S. Torquato, F. H. Stillinger, and R. Connelly, J. Appl. Phys. \textbf{95}, 989 (2004).
\bibitem{st_bc_inprep} S. Tewari and B. Chakraborty, in preparation.
\bibitem{gran_temp_ref} I. Goldhirsch, Ann. Rev. Fluid Mech. \textbf{35}, 267 (2003).
\bibitem{Green-Kubo} Denis J. Evans and Gary Morriss, {\it Statistical Mechanics of Nonequilibrium Liquids}, (Cambridge University Press, 2008), Chapter 4.
\bibitem{SSFR}D. J. Evans and D. J. Searles, Adv. Phys. \textbf{51}, 1529 (2002); G. Gallavotti, Eur. Phys. J. \textbf{B 61}, 1 (2008); U. Seifert, Eur. Phys. J. B 64, 423 (2008).
\bibitem{bray_persistence}S. N. Majumdar, C. Sire, A. J. Bray, and S. J. Cornell, Phys Rev Lett \textbf{77}, 2867 (1996).
\bibitem{garrahan_chandler} J. P. Garrahan and D. Chandler, Phys. Rev. Lett. \textbf{89}, 035704 (2002).
\bibitem{janda} A. Janda, R. Harich, I. Zuriguel, D. Maza, P. Cixous, and A. Garcimart\'{\i}n, Phys. Rev. E. \textbf{79}, 031302 (2009).


\bibitem{drozd_pre08} J. J. Drozd and C. Denniston, Phys Rev E \textbf{78}, 041304 (2008).
\bibitem{gumbel} E. J. Gumbel, {\it Statistics of Extremes} (Columbia University Press, New York,1958).
\bibitem{brey} J.J. Brey, M. I. Garc\'{\i}a de Soria, P. Maynar, and M. J. Ruiz-Montero, Phys. Rev. Lett. \textbf{94}, 098001 (2005).
\bibitem{mounier} A. Mounier and A. Naert, arXiv1208.4029v1 (2012).
\bibitem{sood} S. Majumdar and A. K. Sood, Phys. Rev. E \textbf{85}, 041404 (2012).
\bibitem{nitin} N. Kumar, S. Ramaswamy and A. K. Sood, Phys. Rev. Lett. \textbf{106}, 11801 (2011).
\bibitem{md_inprep} M. Dichter, S. Tewari and B. Chakraborty, in preparation.
\bibitem{ertas_halsey} D. Erta\c{s} and T. C. Halsey, Europhys. Lett. \textbf{60}, 931 (2002).
\bibitem{Silbert_Grest} L. E. Silbert, D. Erta\c{s}, G. S. Grest, T. C. Halsey, and D. Levine, Phys. Rev. E \textbf{65}, 031304 (2002).


	

\end{thebibliography}
\end{document}